\documentclass[12pt,a4paper]{article}
\usepackage[utf8]{inputenc}
\usepackage{amsmath}
\usepackage{amssymb}
\usepackage{graphicx}
\usepackage{booktabs}
\usepackage{subfig}
\usepackage{longtable}
\usepackage{amssymb}
\usepackage{array}

\usepackage{geometry}
\usepackage[
backend=biber,
style=numeric,
sorting=ynt
]{biblatex}

\newcolumntype{C}[1]{>{\centering\arraybackslash}p{#1}}

\newcommand{\vv}{\mathbf{v}}
\newcommand{\ee}{\mathbf{e}}
\newcommand{\rr}{\mathbf{r}}
\newcommand{\mm}{\mathbf{m}}
\newcommand{\xx}{\mathbf{x}}
\newcommand{\eps}{\boldsymbol{\epsilon}}
\newcommand{\sig}{\boldsymbol{\sigma}}

\title{Orb: A Fast, Scalable Neural Network Potential}
\author{Mark Neumann\footnotemark[1], James Gin\footnotemark[1], Benjamin Rhodes\footnotemark[1], Steven Bennett\footnotemark[1],\\  Zhiyi Li\footnote{equal contribution}, Hitarth Choubisa, Arthur Hussey, Jonathan Godwin \\
\\
\texttt{\{mark,jonathan\}@orbitalmaterials.com} \\
Orbital Materials}
\date{\today}

\begin{document}

\maketitle

\begin{abstract}
We introduce Orb, a family of universal interatomic potentials for atomistic modelling of materials. Orb models are 3-6 times faster than existing universal potentials, stable under simulation for a range of out of distribution materials and, upon release, represented a 31\% reduction in error over other methods on the Matbench Discovery benchmark. We explore several aspects of foundation model development for materials, with a focus on diffusion pretraining. We evaluate Orb as a model for geometry optimization, Monte Carlo and molecular dynamics simulations.

\end{abstract}

\section{Introduction}

The design of new functional materials has been a critical part of emerging technologies over the past century. Advancements in energy storage, drug delivery, solar energy, filtration, carbon capture and semiconductors have accelerated due to the discovery of entire classes of materials with application specific properties, such as Perovskites and metal-organic frameworks (MOFs).  However, \textit{ab initio} computational methods \cite{kohn_sham} for designing new inorganic materials are slow and scale poorly to realistically sized systems. New methods using deep learning offer a way to achieve \textit{ab initio} accuracy with dramatically increased speed and scalability.
\paragraph{}

In recent years, deep learning methods have demonstrated their ability to approximate extremely complex natural distributions across a diverse range of application areas including vision, biology and spatial processing, by focusing on architectures that are embarrassingly parallel and can be run efficiently on modern hardware \cite{vaswani2023attentionneed, gnns}, despite lacking architectural biases which would suit the target domain. Our key innovation in Orb is to adapt this approach to materials modeling, focusing on a scalable graph neural network architecture which learns the complexity of atomic interactions and their invariances from data, rather than using architecturally constrained models which respect rotational equivariance or particular group symmetries.

\paragraph{}
Using these methods we have developed Orb, a suite of pre-trained Machine Learning Force Fields (MLFFs) designed as general purpose, universal interatomic potentials. Orb models are available under the Apache 2.0 license, allowing both research and commercial use. Upon release, Orb outperformed all existing methods on the Matbench Discovery Leaderboard, as well as being over 3-6 times faster across a range of commodity hardware. In addition, we have released a model with learned dispersion corrections, making it suitable for accurately modeling materials where Van der Waals forces play a significant role — such as layered materials, molecular crystals, or systems with weak intermolecular interactions.

\section{Model}

The backbone of Orb is a Graph Network Simulator (GNS) \cite{SanchezGonzalez2020LearningTS} augmented with a smoothed graph attention mechanism \cite{Velickovic2017GraphAN}, where messages between nodes are updated based on both attention weights and distance-based cutoff functions. 

\subsection{Graph Construction}
Atomic systems are represented as graphs \(G = (V, E, C)\), where:
\begin{itemize}
    \item \(V = \{\vv_i\}_{i=1}^{N}\) are vector embeddings of the atomic type. Node embeddings explicitly do not contain absolute position information to preserve translational invariance.
    \item \(E = \{\ee_{i, j}\} := \{\big(\frac{\rr_k}{||\rr_{i, j}||}, \ \text{\small{RBF}}(||\rr_{i, j}||)\big)\}\) are the directed edge features, where \(\rr_{i, j}\) is the displacement vector between the \(i^{\text{th}}\) and \(j^{\text{th}}\) atoms. Only atoms within a pbc-aware cutoff-radius from each other are connected, and \(\text{\small{RBF}}: \mathbb{R} \rightarrow \mathbb{R}^{p}\) is a Gaussian Radial Basis Function defined as:
\[
\text{RBF}(x)_l = \exp \left( -\left( \frac{x - \mu_l}{\sigma_l} \right)^2 \right) \hspace{5mm} l \in \{1, \ldots, p\}.
\]
    where \(\mu_l\) represents the center of a Gaussian-shaped function, and \(\sigma_l\) the scale parameter controlling the width.
    \item The edge construction above takes into account periodic boundary conditions according to the optional unit cell \(C \in \mathbb{R}^{3 \times 3}\). We do not provide the unit cell as an input to the model; it is only implicit through the edge construction.
\end{itemize}

\subsection{Model Architecture}
The model is composed of three stages: an \texttt{Encoder}, \texttt{Processor}, and \texttt{Decoder}. When describing these stages, we will make repeated use of multi-layer perceptrons followed by layer normalization \cite{Ba2016LayerN}; we denote this combination by \(\phi\). \\

\noindent The \texttt{Encoder}  computes:
\[
\vv_i^{0} = \phi_{\text{node}}^{0}(\vv_i), 
\quad \ee_{i, j}^0 = \phi_{\text{edge}}^{0}(\ee_{i, j}). \\
\]
\vspace{2mm}
The \texttt{Processor} is a stack of Message Passing layers. Each layer has 4 steps\footnote{For notational simplicity, we omit a layer-indexing \(k\) superscript from some variables.}: \vspace{2mm}
\begin{align*}
\text{(\small{Edge residual})} & & \quad \eps_{i, j} &= \phi_{\text{edge}}^k \left( \left[ \ee_{i, j}^{k}, \, \vv_i^k, \, \vv_j^{k} \right] \right), \\[0.5em]
\text{(\small{Messages})} & & \quad \mm_i^1 &= \sum_{j \in \mathcal{N}_{\text{receive}}(i)} \psi_1^k(\ee_{i, j}^{k}) \, \eps_{i, j}, \quad  \mm_i^2 = \sum_{j \in \mathcal{N}_{\text{send}}(i)} \psi_2^k(\ee_{j, i}^{k}) \, \eps_{j, i} ,\\[0.5em]
\text{(\small{Node update})} & & \quad \vv_i^{k+1} &=  \vv_i^k + \phi_{\text{node}}^{k} \left( \left[ \vv_i^{k}, \, \mm_i^1, \, \mm_i^2  \right] \right), \\[0.5em]
\text{(\small{Edge update})} & & \quad \ee_i^{k+1} &= \ee_i^k + \eps_{i, j} ,
\end{align*}

\vspace{3mm}
\noindent where \(\mathcal{N}_{\text{receive}}(i), \ \mathcal{N}_{\text{send}}(i) \) are index sets for receiving/sending atoms. The messages are weighted sums of edge residuals, where \(\psi_1, \psi_2\) are scalar-valued ``cutoff gates": \vspace{2mm}
\begin{align*}
\hspace{6mm} \psi_m^k(\ee_{i, j}^k) &= \text{\texttt{sigmoid}}(\mathbf{w}_m^k \cdot\ee_{i, j}^k) \cdot \text{\texttt{cutoff}}(||\rr_{i, j}||), \hspace{6mm} m \in \{1, 2\}, \\[0.6em]
    \text{\texttt{cutoff}}(r) &= \left(1 - \left(\frac{r}{r_{\max}}\right)^p\right)_+.
\end{align*}

\vspace{2mm}
\noindent where the purpose of \texttt{cutoff} is to ensure that the model is continuous with respect to the input displacement vectors \(\rr_{i, j}\). Without such a cutoff, small perturbations to these displacements can cause discontinuous jumps in model predictions.
\paragraph{}

The \texttt{Decoder} takes the output of the \texttt{Processor} and predicts the next state of the system by projecting the node representations \(\vv_i\) using 3 independent MLPs:
\begin{itemize}
    \item \(E \in \mathbb{R} = \text{MLP}_{\text{energy}}( \frac{1}{N} \sum_{i} \vv_i)\), the scalar total energy. 
    
    \item \(\mathbf{f}_i \in \mathbb{R}^ = \text{MLP}_{\text{forces}}(\vv_i)\), the per atom force vectors. 
    
    \item 
    \(\sig \in \mathbb{R}^{6} = \text{MLP}_{\text{stress}}(\frac{1}{N} \sum_{i} \vv_i)\), the unit cell stress in reduced Voigt notation. 
\end{itemize}

At inference time, our model requires a single forward pass to produce \( E, \mathbf{f}_i\) and \(\sig\), as we do not implement our forcefield as a conservative vector field. This halves the amount of computation required for computing forces, but does require two adjustments to ensure net zero force and torque. 

These two constraints are as follows:

\begin{itemize}
    \item \(\mathbf{\hat{f}}_i{f}_i - \frac{1}{N} \sum_{i=1}^{N} \mathbf{f}_i\) \hspace{5mm} (\textit{net zero force})
    \item \(\mathbf{\Tilde{f}}_i^{\text{pred}} = \mathbf{\hat{f}}_i + \delta \mathbf{f}_i\) \hspace{13mm} (\textit{net torque removal})
\end{itemize}

where $\delta \mathbf{f}_i$ is the smallest (in L2 norm) additive adjustment to the forces whose net torque cancels with the net torque of \(\mathbf{\hat{f}}_i\), whilst maintaining the zero net force property. This adjustment is only applied to non-pbc systems, and is implemented by minimizing a Lagrangian incorporating these two constraints.  See Appendix \ref{section:torque_removal} for a full description of these adjustments.

\subsection{Training}

Orb models are trained in two distinct phases - first they are trained as Denoising Diffusion models \cite{sohl2015deep, Song2019GenerativeMB, Ho2020DenoisingDP} using a dataset of ground state materials. Secondly, the base models are used to initialize a Neural Network Potential which predicts the energy, forces and stress of systems and is trained in a supervised manner.

\subsubsection{Diffusion Model Pretraining}
Given initial ground state atomic positions \(\xx_0\), a forward diffusion process gradually adds noise to the system. The noisy positions at time step \(t\), denoted \(\xx_t\), are generated by a conditional Gaussian
\[
q(\xx_t | \xx_0) = \mathcal{N}(\xx_t ; \  \xx_0, \sigma_t \mathbf{I}),
\]
where \(\sigma_t\) is a noise scale that increases over time, and \(\mathbf{I}\) is the identity matrix.

We then estimate a sequence of score functions of the form \( \mathbf{s}_t(\xx) = \nabla_{\xx} \log p_t(\xx) \) by minimizing a multi-scale denoising objective \cite{hyvarinen2005estimation, vincent2011connection, Song2019GenerativeMB}:
\[
\sum_{t=0}^{T} \lambda(t) \ \ \mathbb{E}_{q(\xx_t|\xx_0)} \  || s_{\boldsymbol{\theta}}(\xx_t, t) - \nabla_{\xx} \log q(\xx_t | \xx_0) ||^2 \hspace{7mm}
\]
which, following a standard reparameterization, gives an ``epsilon prediction loss"
\[
\sum_{t=0}^{T} \lambda'(t) \ \ \mathbb{E}_{N(\eps ; \  0, 1)} \  \Big|\Big| \frac{s_{\boldsymbol{\theta}}(\xx_t, t)}{\sigma_t} + \eps \Big|\Big| ^2, \hspace{7mm} \text{where} \ \  \xx_t = \xx_0 + \sigma_t \eps. \hspace{7mm}
\]
\paragraph{}

\subsubsection{NNP Finetuning} 
\label{sec:nnp-finetune}

The base diffusion model is finetuned on energies, forces and stresses of DFT optimization trajectories. The energy loss, \(\mathcal{L}_{E}\), per datapoint has the form
\[
\mathcal{L}_{E} = \left|(E_{\text{pred}} - \frac{(E_{\text{true}} - E_{\text{ref}})}{N} \right| ,
\]
where \(E_{\text{pred}}\) is the output of the MLP energy head, \(E_{\text{true}}\) is the true total energy, and \(E_{\text{ref}}\) is a fixed, linear reference energy based on the stoichiometry of the system.

The forces loss \(\mathcal{L}_{\mathbf{f}}\) is then computed as the MAE between predicted forces and the true forces:
\[
\mathcal{L}_{\mathbf{f}} = \frac{1}{3N} \sum_{i=1}^{N} \left\| \mathbf{f}_i^{\text{pred}} - \mathbf{f}_i^{\text{true}} \right\|_1
\]
For batched systems, this equation still holds, but \(\mathbf{f}_i^{\text{pred}}, \ \mathbf{f}_i^{\text{true}}\) are concatenated vectors, and \(N\) equals the number of atoms in the batch. In other words, we use a single average over all atoms in the batch, rather than a nested average.

The total loss \(\mathcal{L}_{\text{total}}\) is a weighted sum of the energy, forces, and unaltered stress MAE . The energy loss is optionally scaled by a weight \(\lambda_E\) to balance its contribution:
\[
\mathcal{L}_{\text{total}} = \lambda_E \mathcal{L}_{E} + \mathcal{L}_{\mathbf{f}} + \mathcal{L}_{\sigma}.
\]

\subsubsection{Datasets}

For pretraining, we curate a dataset of minimum energy configurations of materials, regardless of exchange-correlation functional, DFT software implementation or correction. Our motivation for pretraining is to provide an approximate base model which has broad materials coverage in terms of atomic type, materials class, symmetry groups and usage domain. Combining disparate datasets in this way is possible because all we require are atomic positions and unit cells; all other properties and metadata are ignored, and it is usually these properties (such as potential energy) that are incompatible across datasets.
\paragraph{}

In comparison, for finetuning a UIP, data quality and consistency is of critical importance. Here, we follow previous work \cite{zeni2024mattergengenerativemodelinorganic, GNOMe} in combining MPtraj, the dataset constructed from the Materials Project and used to train CHGNet \cite{deng2023chgnetpretraineduniversalneural}, with Alexandria \cite{alexandria}, as both use PBE exchange-correlation functionals; Hubbard U corrections for transition metal oxides; no additional dispersion correction; and the same DFT software implementation (VASP) for geometry optimization. MatterGen and MatterSim \cite{yang2024mattersim, zeni2024mattergengenerativemodelinorganic} additionally make use of ICSD, which we disregard as it requires a commercial license. GNOMe \cite{GNOMe} use a large proprietary dataset of geometry relaxations derived from the Materials Project and OQMD \cite{OQMD} via a variant of  \emph{ab initio} random structure search (AIRSS) \cite{abinitio_random_struct_search}.

\begin{figure}
    \centering
    \includegraphics[width=0.9\linewidth]{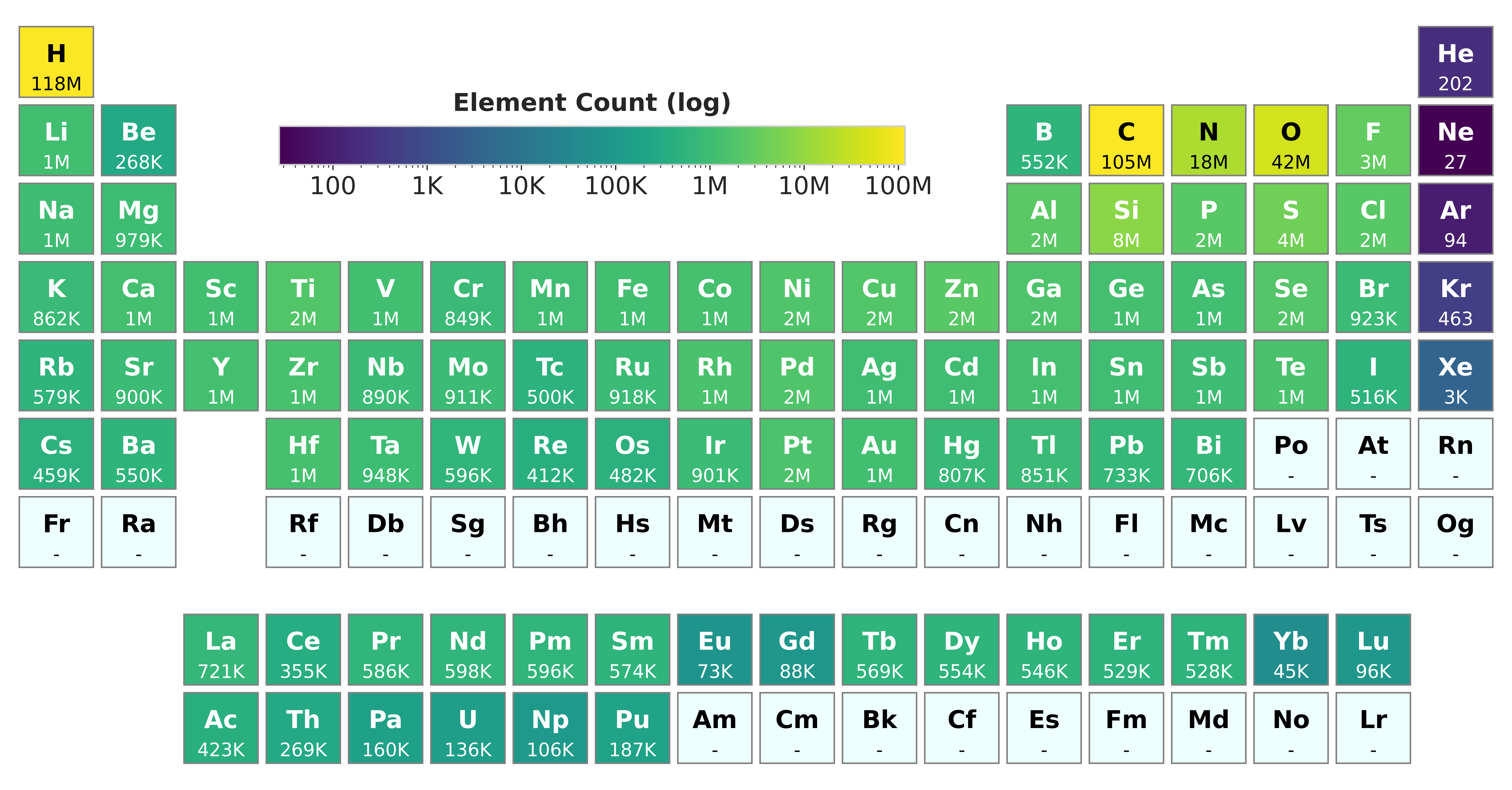}
    \includegraphics[width=0.9\linewidth]{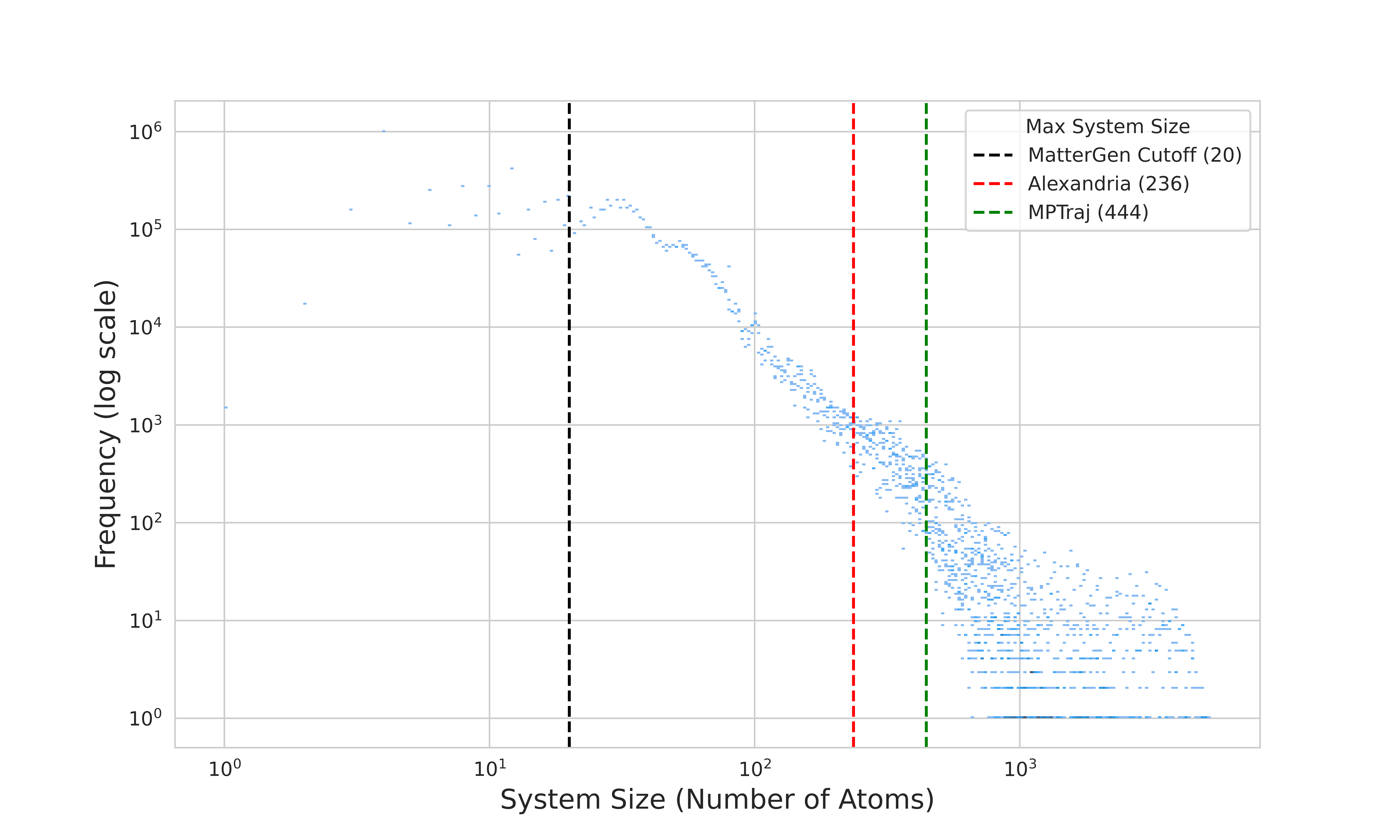}

    \caption{\textit{top}: Element distribution of the full diffusion pretraining dataset for Orb. \textit{bottom}: Distribution of the dataset with respect to system size. Due to the efficiency of Orb's architecture, we can train on systems with up to 5000 atoms, two orders of magnitude larger than MatterGen \cite{zeni2024mattergengenerativemodelinorganic}, a similar generative model of materials.}
    \label{fig:dataset_distribution}
\end{figure}

\subsection{Dispersion Corrections}

Both the MPtraj and Alexandria datasets are generated using DFT with the Perdew Burke Ernzerhof (PBE) exchange-correlation functional \cite{Perdew1996GeneralizedGA}. PBE is a semi-local functional which does not capture long range dispersion interactions. To account for long range intermolecular interactions, additional corrections such as D3 can be applied \cite{Grimme2010ACA}. Because D3 is an additive correction, it is possible to apply it \emph{at inference time} to any existing UIP trained on PBE data, as demonstrated by \cite{batatia2024foundationmodelatomisticmaterials}. However, this causes significant slowdowns in practice, given D3's $n^2$ complexity with respect to system size (see Section \ref{section:speed}).
\paragraph{}
We amortize this cost by adding D3 corrections to the MPtraj and Alexandria datasets and then training Orb-D3 models on this corrected data. We note that the receptive field of Orb is more than sufficient to model such long range dispersion interactions. Unless specified, model comparisons in the paper \textbf{do not} use Orb-D3, so as to remain comparable to existing UIPs such as MACE.

\section{Results}
\subsection{Matbench-Discovery}
\begin{table}[!t]
    \centering
\resizebox{\textwidth}{!}{%
\begingroup
\begin{tabular}{lccccccc}
    \toprule
    Model & F1 & DAF & Precision & Recall & Accuracy & MAE  & R2 \\
    \midrule
    Voronoi RF & 0.344 & 1.509 & 0.259 & 0.511 & 0.665 & 0.142 & -0.314 \\
    BOWSR & 0.437 & 1.836 & 0.315 & 0.711 & 0.702 & 0.115 & 0.141 \\
    Wrenformer & 0.479 & 2.13 & 0.365 & 0.693 & 0.741 & 0.105 & -0.036 \\
    CGCNN+P & 0.51 & 2.398 & 0.411 & 0.67 & 0.779 & 0.108 & 0.026 \\
    CGCNN & 0.51 & 2.631 & 0.451 & 0.587 & 0.807 & 0.135 & -0.62 \\
    MEGNet & 0.513 & 2.699 & 0.463 & 0.574 & 0.813 & 0.128 & -0.276 \\
    ALIGNN & 0.565 & 2.921 & 0.501 & 0.649 & 0.829 & 0.092 & 0.273 \\
    M3GNet & 0.576 & 2.647 & 0.454 & 0.788 & 0.802 & 0.072 & 0.584 \\
    CHGNet & 0.612 & 3.038 & 0.521 & 0.74 & 0.839 & 0.061 & 0.685 \\    
    MACE & 0.668 & 3.4 & 0.583 & 0.781 & 0.867 & 0.055 & 0.698 \\
    SevenNet & 0.724 & 4.252 & 0.650 & - & 0.904 & 0.048 & 0.750 \\
    Orb (MP-trj Only) & 0.764 & 4.695 & 0.718 & 0.818 & 0.923 & 0.045 & 0.756 \\
    \midrule
    GNoME* & 0.81 & 4.81 & 0.825 & 0.796 & 0.942 & 0.034 & 0.781 \\
    MatterSim* & 0.832 & 4.838 & 0.83 & 0.834 & 0.942 & 0.026 & 0.804 \\
    Orb & 0.880 & 6.035 & 0.923 & 0.842 & 0.965 & 0.028 & 0.824 \\
    \footnotesize{no protostructure overlap}
     & 0.877 & 6.01 & 0.919 & 0.839 & 0.964 & 0.029 & 0.823 \\
    \bottomrule
\end{tabular}
\endgroup
}
    \caption{Matbench Discovery results using Orb. Structures are relaxed using Orb and the FIRE \cite{Bitzek2006StructuralRM} optimizer for 500 steps or until the maximum force magnitude is lower than 0.05 eV/\AA. Models marked with an asterisk are closed source.}
    \label{table:matbench}
\end{table}

We evaluate Orb on the widely adopted Matbench-Discovery benchmark \cite{riebesell2024matbenchdiscoveryframework}. This is a high-throughput geometry optimization and stability prediction benchmark that demands accurate estimates of energies, forces, and stress tensors at \emph{ab initio} fidelity. To succeed, a model must relax a large test set of crystal structures to a ground-state, and then predict the decomposition energy of this ground-state relative to a convex hull of competing phases. Accurately determining the energy above or below the convex hull with DFT-level precision has experimental implications, serving as an indicator for a material’s synthesizability.
\paragraph{}

The full results, accurate as of the time of submission (2024-10-14), are in Table \ref{table:matbench}. Models in the top section are MPtraj-only, whilst models in the second section are unrestricted. For both sections, Orb models set a new state-of-the art across the primary metric (F1), and are generally superior across other metrics. Orb models have especially high precision, which would likely be crucial in a real-world use-case where false-positives (i.e.\ materials that are predicted as stable but are not) are very costly due to wasted experimental effort. The final row in Table \ref{table:matbench} refers to a version of Orb trained on a reduced training set. Specifically, we removed all datapoints from Alexandria that had overlapping protostructures with the WBM test set. We use the same approach as \cite{riebesell2024matbenchdiscoveryframework} to identify protostructrures with \texttt{aviary.wren.utils.get\_protostructure\_label\_from\_spglib}\footnote{Using \texttt{v1.1.0} of the aviary pypi package.} See Appendix \ref{appendix:omat} for discussion of the concomitantly released OMAT24 dataset and models \cite{barrosoluque2024openmaterials2024omat24} which outperform Orb-v2 on Matbench Discovery.

\subsection{Effect of Pretraining}

\begin{figure}
    \centering
    \hspace*{-0.9cm}
    \includegraphics[width=1.1\linewidth]{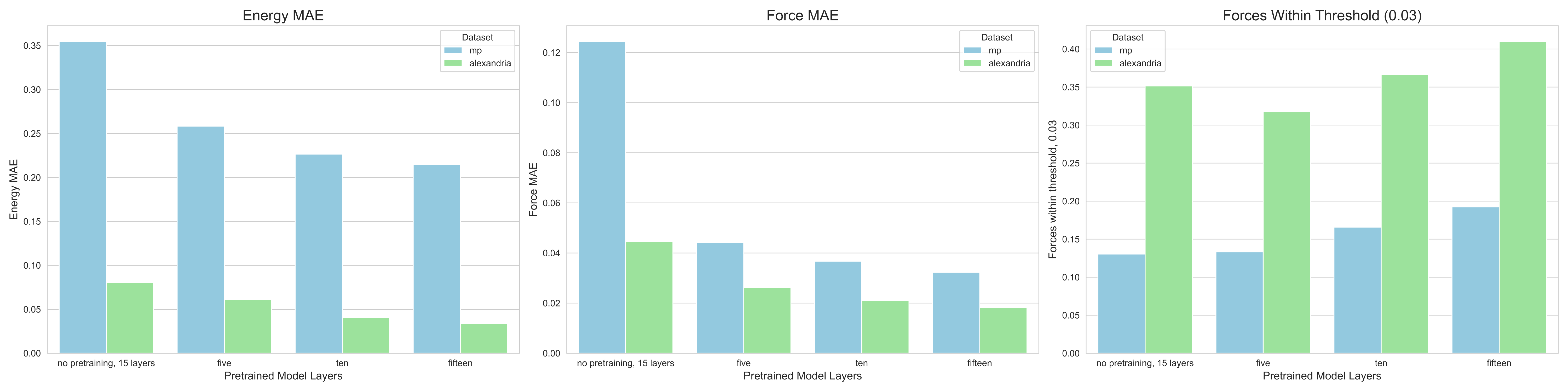}

    \caption{Effect of pretraining and model size on Alexandria and MPtraj Energy MAE, Forces MAE and Forces within a threshold of 0.03. Diffusion pretraining helps universally with improvements between 17\% and 70\%, even on Alexandria, which is an order of magnitude larger MPtraj.}
    \label{fig:ablation}
\end{figure}

Figure \ref{fig:ablation} shows the effect of our pretraining procedure on the MP-traj and Alexandria validation metrics. In addition to the substantial decreases in forcefield error, diffusion pretraining also makes training more stable; anecdotally we have observed fewer instances of oversmoothing \cite{li2018deeperinsightsgraphconvolutional} when finetuning diffusion pretrained models compared to training forcefields from scratch. 
\paragraph{}
Although correlated, improvements from pretraining are not only related to the amount of data seen during pretraining - pretraining on existing datasets also results in substantially improved performance, as evidenced by the \texttt{Orb-v2-mptraj} performance on the Matbench Discovery Benchmark (Table \ref{table:matbench}).

\subsection{Speed Benchmark}
\label{section:speed}

\begin{figure}
    \includegraphics[width=0.49\linewidth]{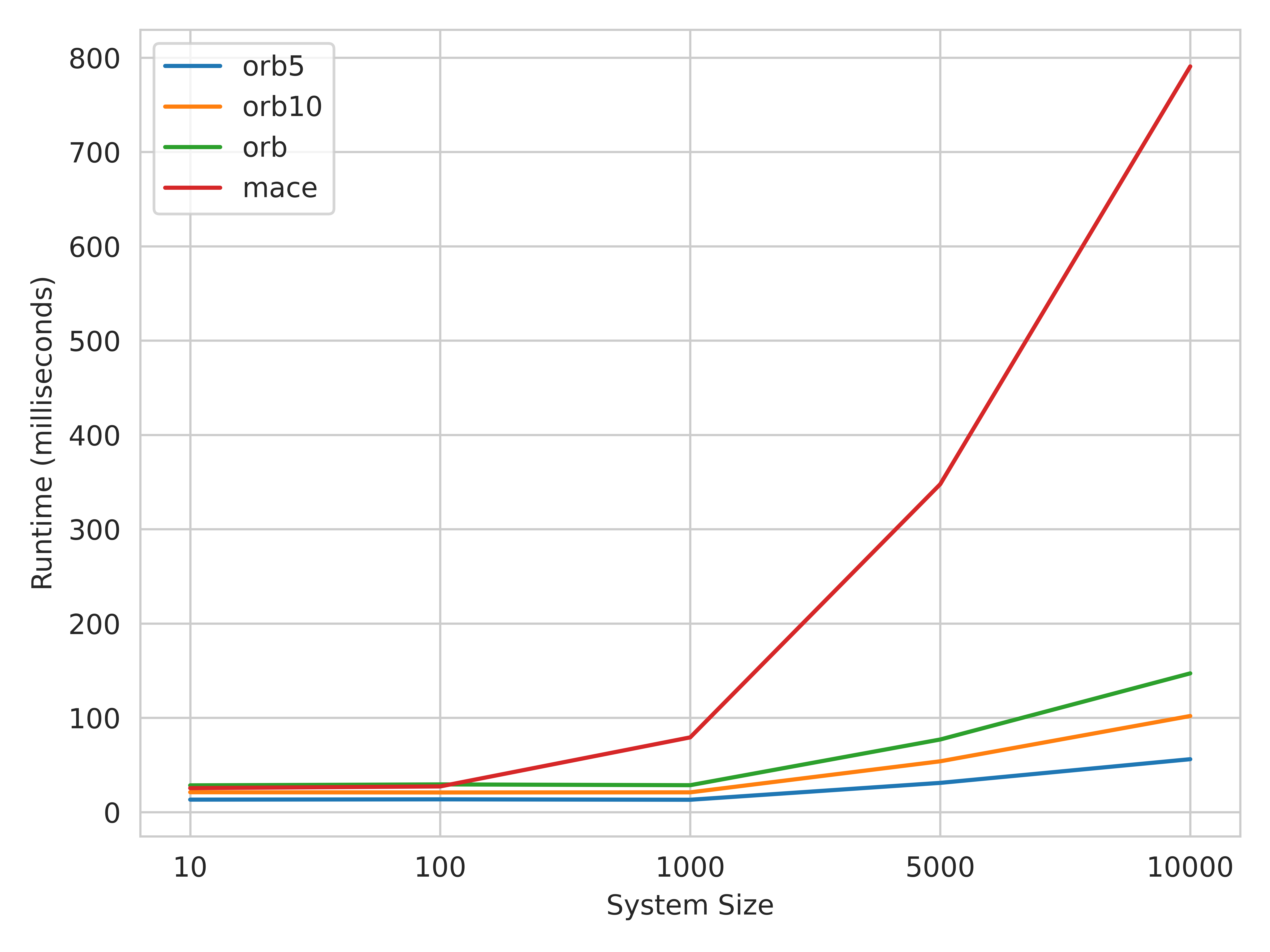}\hfill
    \includegraphics[width=0.49\linewidth]{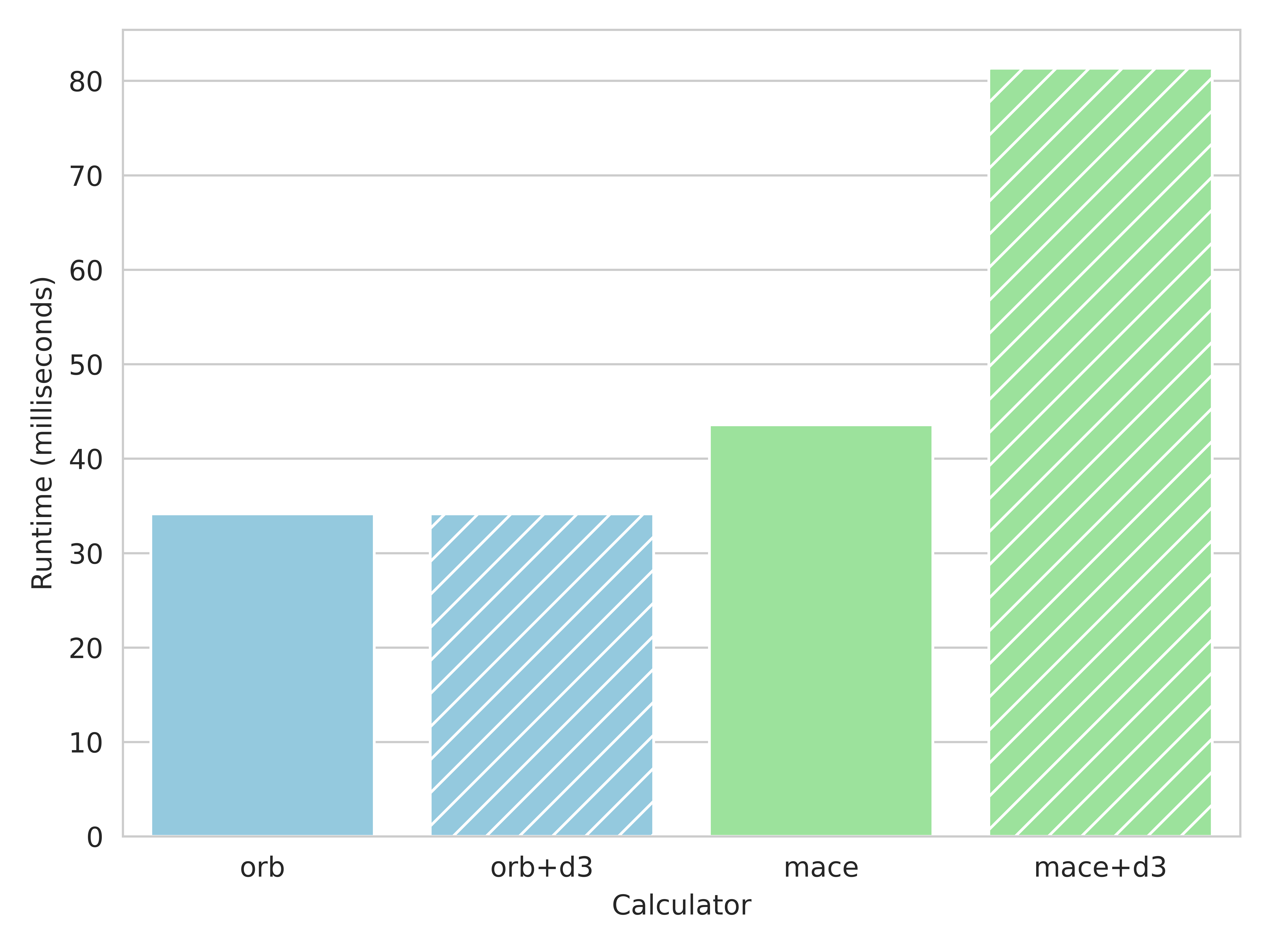}

    \caption{\textit{left:} Model forward pass speed (excluding featurization) compared to MACE on a single NVIDIA A100 GPU. At large system sizes, Orb is between 3 to 6 times faster than MACE. \textit{right:} End to end model inference speed for a 100 atom system on a single NVIDIA A100 when implemented as a \texttt{Calculator} object in the Atomic Simulation Environment Python library. The D3 dispersion correction adds a substantial cost which is amortized by Orb models, as the corrections are incorporated into training datasets. All measurements reported as the median of 50 runs.}
    \label{fig:speed_benchmark}
\end{figure}

In addition to producing highly accurate UIPs for both optimization and simulation, Orb models are considerably faster and more scalable than existing open source neural network potentials. Figure \ref{fig:speed_benchmark} shows relative performance improvements compared to MACE, particularly at large system sizes. Efficient scaling with respect to system size is essential to model certain phenomena such as dopants, or to accurately capture sparse simulation statistics such as diffusivity. 

\subsection{Molecular Dynamics of Small Molecules}
\label{sec:md17-section}

We evaluate Orb on the MD17-10k benchmark, following the protocol of \cite{fu2022forces}. This benchmark assesses the quality of NVT Molecular Dynamics (MD) trajectories generated by machine learning models on four small organic molecules (aspirin, ethanol, naphthalene, and salicylic acid) using two key metrics: 
\begin{itemize}
    \item Stability: the number of picoseconds until the system ``collapses" / ``breaks apart", which is detected via large deviations in the interatomic distances compared to those of a ground-truth \emph{ab initio} trajectory. 300 ps is the length of the simulation and thus the maximum possible stability.
    \item ``h(r)": a 1-Wasserstein metric on the distributions of interatomic distances present in the model-trajectory vs those present in the \emph{ab initio} trajectory. This metric is similar in spirit to a radial distribution function error metric.
\end{itemize}
In addition to these two metrics, we report a test set Force MAE (meV/\r{A}). This MAE is not of interest by itself (since it is not a function of the MD simulation) but it is interesting to correlate it against MD performance.

\subsubsection*{MD17-10k: molecule specific results}
\label{sec:md17-molecule-specific-models}
\begin{table}[!t]
    \centering
\hspace*{-0.9cm}
\resizebox{1.1\textwidth}{!}{%
\begingroup
\setlength{\tabcolsep}{2.5pt} 
\begin{tabular}{llccccccccc}
\toprule
    Metric & Molecule & DeepPot-SE & SchNet & DimeNet & PaiNN & SphereNet & ForceNet & GemNet-dT & NequIP & Orb \\
\midrule
    Stability ($\uparrow$) & \small{Aspirin}      & 9     & 26    & 54    & 159   & 141   & 182   & 192   & 300   & 300 \\
                  & \small{Ethanol}      & 300   & 247   & 26    & 86    & 33    & 300   & 300   & 300   & 300 \\
                  & \small{Naphthalene}  & 246   & 18    & 85    & 300   & 6     & 300   & 25    & 300   & 300 \\
                  & \small{Salicylic Acid} & 300 & 300   & 73    & 281   & 36    & 1     & 94    & 300   & 300 \\
\cmidrule{2-11}
                  & \small{AVG}          & 213.8& 147.8& 60.0  &206.5  & 54    &195.8 &152.8 & 300   & 300 \\
\midrule
    $h(r)$ ($\downarrow$)    & \small{Aspirin}      & 0.65  & 0.36  & 0.04  & 0.04  & 0.03  & 0.56  & 0.04  & 0.02  & 0.03 \\
                  & \small{Ethanol}      & 0.09  & 0.21  & 0.15  & 0.15  & 0.13  & 0.86  & 0.09  & 0.08  & 0.09 \\
                  & \small{Naphthalene}  & 0.11  & 0.09  & 0.10  & 0.13  & 0.14  & 1.02  & 0.12  & 0.12  & 0.12 \\
                  & \small{Salicylic Acid} &0.03  & 0.03  & 0.06  & 0.03  & 0.06  & 0.35  & 0.07  & 0.03  & 0.03 \\
\cmidrule{2-11}
                  & \small{AVG}          & 0.22  &0.17 &0.088 &0.088 & 0.090  &0.70 &0.080  &0.063 &0.068 \\
\midrule
    Force ($\downarrow$) & \small{Aspirin}         & 21.0    & 35.6  & 10.0    & 9.2   & 3.4   & 22.1  & 5.1   & 2.3   & 2.4 \\
               & \small{Ethanol}         & 8.9   & 16.8  & 4.2   & 5.0     & 1.7   & 14.9  & 1.7   & 1.3   & 1.7 \\
               & \small{Naphthalene}     & 13.4  & 22.5  & 5.7   & 3.8   & 1.5   & 9.9   & 1.9   & 1.1   & 1.0   \\
               & \small{Salicylic Acid}  & 14.9  & 26.3  & 9.6   & 6.5   & 2.6   & 12.8  & 4.0     & 1.6   & 1.8 \\
\cmidrule{2-11}
               & \small{AVG}             & 14.6 & 25.3  & 7.4 & 6.1 & 2.3   & 15.0 & 3.2 & 1.6 & 1.7 \\
\bottomrule
\end{tabular}
\endgroup
}

    \caption{Results on the MD17 Molecular Dynamics benchmark. With the exception of Orb, all results are taken from \cite{fu2022forces}. Stability is measured in picoseconds, with large values meaning more stable, and 300 being the maximum possible value. For $h(r)$ and Force, lower values are better and 0 is the minimum.}
    \label{table:md17}
\end{table}
Most prior work using NNPs for molecular simulation train \emph{system specific} models. The MD17-10K benchmark is no exception, with all models reported in \cite{fu2022forces} being trained from scratch per individual system (e.g.\ aspirin), with no expectation that the model could simulate any other small organic molecule.
\paragraph{}
In order to compare with this prior work, we also train system specific models by finetuning four versions of Orb, one for each molecule (Table \ref{table:md17}). Orb demonstrates excellent stability, achieving the maximum 300ps for all molecules, putting it on par with NequIP and ahead of all other baseline models. Orb also performs well in terms of the \emph{realism} of its simulated trajectories, closely matching the reference data as shown by its low \emph{h(r)}, again competitive with NequIP.

\subsubsection*{MD17-10k: zero-shot results}
\label{sec:md17-zero-shot-models}

Training system specific models is expensive and unscalable. It is therefore of interest to directly evaluate Orb without additional finetuning. Such a zero-shot evaluation is quite ambitious; ow well can models trained on \emph{zero-kelvin periodic crystal structures} generalize to unseen \emph{high-temperature non-periodic molecules}?"
\paragraph{}

The results for Orb and several other universal NNPs are shown in Table \ref{table:md17-zeroshot}. We present only $h(r)$ as all simulations are stable and Force MAE is not computable due to changes in the level of theory for the DFT computation. Comparing just those models trained on MPtraj, we see that Orb has the best performance, with a lower $h(r)$ across the board. It is noteworthy that Orb (MPtraj) outperforms many of the molecule-specific baselines in Table \ref{table:md17}, and is not far from matching the molecule-specific versions of Orb.
\paragraph{}
Whilst Orb (MPtraj) does outperform MACE and Sevennet (Table \ref{table:md17-zeroshot}), all models are reasonably close in performance and it is hard to make a robust judgement of superiority from a sample size of four simulations. Indeed, the somewhat surprising result that adding in Alexandria to Orb's training set lowers performance may not be statistically significant across a larger sample of molecules.
\paragraph{}
\begin{table}[!t]
    \centering
\hspace*{-0.9cm}
\resizebox{1.1\textwidth}{!}{%
\begingroup
\setlength{\tabcolsep}{2.5pt} 
\begin{tabular}{llccccc}
\toprule
    \small{Metric} & \small{Molecule} & \parbox{3cm}{\centering \small{MACE} \\ \footnotesize{(MPtraj)}} & \parbox{3cm}{\centering \small{SevenNet} \\ \footnotesize{(MPtraj)}} & \parbox{3cm}{\centering \small{Orb} \\ \footnotesize{(MPtraj)}} & \parbox{3cm}{\centering \small{Orb} \\ \footnotesize{(MPtraj+Alex)}} & \parbox{3cm}{\centering \small{Orb w/o NTR} \\ \footnotesize{(MPtraj+Alex)}} \\
\midrule
    $h(r)$ ($\downarrow$)  & \small{Aspirin}        & 0.06	& 0.07 & 0.06 & 0.11 & 0.37  \\
                & \small{Ethanol}        & 0.12	& 0.13 & 0.11 & 0.1 & 0.16  \\
                & \small{Naphthalene}    & 0.15 & 0.12 & 0.11 & 0.1 & 0.11  \\
                & \small{Salicylic Acid} & 0.09	& 0.06 & 0.05 & 0.04 & 0.21 \\
\cmidrule{2-7}
                  & \small{AVG}          & 0.105 & 0.095 & 0.0825 & 0.0875 & 0.2125 \\
\bottomrule
\end{tabular}
\endgroup
}
    \caption{Results on our newly introduced zero-shot version of the MD17 Molecular Dynamics benchmark. All models shown here were trained on \emph{zero-kelvin periodic crystal structures} and must extrapolate at test-time to 500 Kelvin isolated molecules with no further training or finetuning on individual trajectories. NTR = Net Torque Removal; see Section \ref{sec:nnp-finetune} for explanation.}
    \label{table:md17-zeroshot}
\end{table}

However, we do believe an important conclusion can be drawn from the final column in Table \ref{table:md17-zeroshot}, which shows an ablation of Orb without the ``Net Torque Removal'' (NTR) operation described in Section \ref{section:torque_removal}. This operation is an inexpensive ``correction" to the model's predicted forces that removes net torque when processing non-periodic systems. Intriguingly, this operation can be applied purely at inference-time, with minimal effect on Force MAEs, but a substantial effect on the quality of MD simulations. 
\paragraph{}

The need for this operation---as well as the `mean removal' operation also described in Section \ref{section:torque_removal}---arises from the fact that Orb is \emph{non-conservative} i.e. it directly predicts forces rather than computing the gradient of an energy function. In contrast, conservative models like MACE automatically generate force predictions that have zero net force and (for non-periodic systems) zero net torque\footnote{Strictly speaking being conservative is not sufficient for these properties; zero net force also requires translation invariance and zero net torque requires rotational equivariance.}. Our conclusion is that computing gradients of energy functions is not intrinsically necessary for stable MD; it is just one strategy to ensure certain \emph{desirable properties} for the force predictions. These properties can be acquired in other ways, enabling non-conservative potentials to perform equally well, if not better.

\subsection{Molecular Dynamics of Crystalline Materials}

In addition to evaluating the performance of Orb on isolated small molecules, we extend our evaluation to the stability of MD simulations at increasingly higher temperatures on larger bulk crystalline materials.
We evaluate the MD stability of the well-studied MOF-5, recorded to be stable up to temperatures of 723 K\cite{yang2011methyl}, making it an ideal candidate to test Orb's simulation capabilities under thermal stress.

\begin{figure}[!t]
    \centering
    \includegraphics[width=\linewidth]{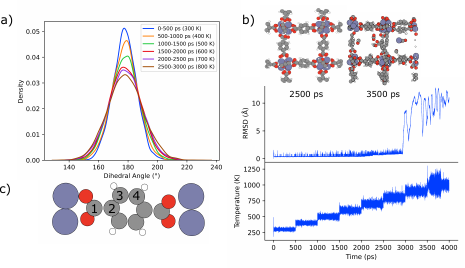}
    \caption{(a) Distribution of dihedral angles across the MD trajectory for MOF-5, shown at six time intervals (0–500, 500–1000, 1000–1500, 1500–2000, 2000–2500, and 2500–3000 ps) corresponding to increasing temperatures from 300 K to 800 K; (b) \textit{top}: snapshots of MOF-5 at 2,500 ps (stable framework) and 3,500 ps (decomposed structure), \textit{middle}: root mean squared deviation (RMSD) from the initial structure, remaining constant until around 3,000 ps, after which significant structural degradation occurs, \textit{bottom}: simulation temperature over time, showing incremental heating; (c) schematic of the four carbon atoms ($C1$–$C4$) in the MOF-5 and linker structure used to calculate the dihedral angle.}
    \label{fig:mof-md}
\end{figure}

\paragraph{}
Here, we show that Orb produces stable MD trajectories for MOF-5 across a range of different temperatures. Figure \ref{fig:mof-md}b shows the root mean squared deviation (RMSD) from the starting state remains relatively constant until 3,000 ps, corresponding to a temperature of 800 K. Beyond this point, significant degradation of the framework begin to occur into gaseous constituents.
In addition to a stable simulation at a range of temperatures, we show Orb is able to replicate the dynamic behavior of MOF-5 observed experimentally. The interaction of the phenylene linker with neighboring atoms in MOF-5 has been shown to prohibit full rotation of the phenylene ring.\cite{lock2010elucidating} In Figure \ref{fig:mof-md}a, we show the distribution of dihedral angles of MOF and linker atoms $C_1$-$C_2$-$C_3$-$C_4$ (Figure \ref{fig:mof-md}c). We observe no complete rotation of the phenylene ring to occur between 300-800 K. At higher temperatures, however, the broader distribution of dihedral angles indicates a greater degree of rotational flexibility due to increased thermal energy.

\paragraph{}
The observed dynamics show that Orb can be extended beyond small molecule MD to bulk crystalline materials. Despite not being explicitly trained on MOFs, we show stable simulations across a range of temperatures reported to be stable in the literature. 
In addition, we show linker dynamics in agreement with those observed in the literature, suggesting Orb can be used to investigate the mechanical properties of larger crystalline materials.

\subsection{Adsorption}

We evaluate Orb for modeling the dynamic adsorption properties of a prototypical metal-organic framework (MOF) for low pressure  $\text{CO}_2$ adsorption. Mg-MOF-74 is a promising candidate for $\text{CO}_2$ adsorption due to the presence of open metal sites for  $\text{CO}_2$ to adsorb that are accessible via the one-dimensional hexagonal channels.\cite{caskey2008dramatic}

\paragraph{}
To investigate the adsorption properties of Mg-MOF-74 at low pressures, we use Widom insertion using Orb-D3 to compute the heat of adsorption and free energy landscape of $\text{CO}_2$\cite{widom1963some}. Vandenbrande \textit{et al.} used Widom insertion with a classical force field, before using importance sampling for DFT to compute an \textit{ab initio} heat of adsorption, obtaining a value of –34.1 kJ/mol at the PBE+D3(BJ) level of theory.\cite{vandenbrande2018ab} In this evaluation, we sample the full \textit{ab initio} potential energy to compute the heat of adsorption.

\begin{figure}[!t]
    \centering
    \includegraphics[width=\linewidth]{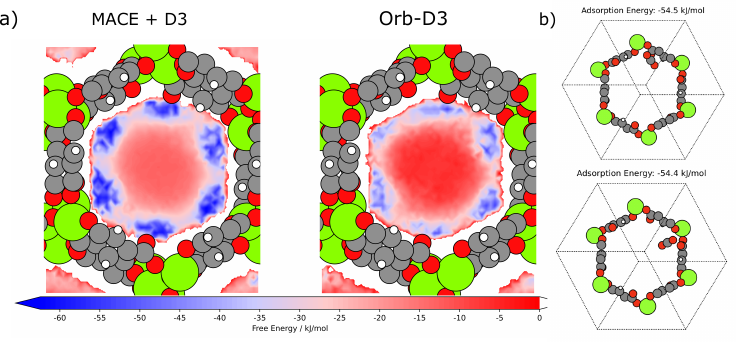}
    \caption{(a) Free energy surface of MACE + D3 (left) and Orb-D3 (right) obtained from Widom insertion in Mg-MOF-74. Lowest free energies are represented by the blue region near the open-metal centers, indicating favorable adsorption sites for $\text{CO}_2$. (b) $\text{CO}_2$ adsorption locations in Mg-MOF-74 showing the two sites with the most favorable adsorption energies, obtained with Widom insertion, with values of -54.5 kJ/mol and -54.4 kJ/mol for the respective sites. While both Orb and MACE predict similar locations of energy minima, the free energy minima of ORB are closer in magnitude to the experimental heat of adsorption (-44 kJ/mol).}
    \label{fig:mof-adsorption}
\end{figure}

\paragraph{}
Figure \ref{fig:mof-adsorption}(a) shows the free energy profile of Mg-MOF-74 computed through Widom insertion with MACE + D3 and Orb-D3. Both MACE + D3 and Orb-D3 show clusters of favorable adsorption sites next to the metal center, indicated by the negative free energy. While both MACE + D3 and Orb-D3 predict similar adsorption regions, there are differences in the depth of the potential wells, with MACE's wells being noticeably deeper. This extra depth leads to a low heat of adsorption (-61.7 kJ/mol) compared to the experimental value of -44 kJ/mol\cite{queen2014comprehensive}. In contrast, Orb-D3's shallower wells closer to the experimental value, with a heat of adsorption of -54.9 kJ/mol.

\paragraph{}
For the energy adsorption site identified by Orb using Figure \ref{fig:mof-adsorption}(b), $\text{CO}_2$ adopts a tilted configuration with a Mg-O-C bond angle of 134.7$^{\circ}$ and a Mg-O($\text{CO}_2$) bond length of 2.43 Å. This geometry is close to the experimentally determined orientation of $\text{CO}_2$ with an experimental Mg-O($\text{CO}_2$) distance of 2.27 Å and Mg-O-C bond angle of 131$^{\circ}$, suggesting reasonable agreement with the experimental binding site and in agreement with the binding site identified in MACE.\cite{batatia2024foundationmodelatomisticmaterials}

\paragraph{}
As neither MACE or Orb are trained on MOF structures, the qualitatively correct energy landscape and adsorption dynamics suggest both Orb and MACE UIPs are able to generalize to complex porous materials. The higher interaction energies and heats of adsorption suggest that further fine-tuning data may be required to get quantitatively correct heats of adsorption for sampling the potential energy landscape using Monte Carlo methods, but both serve as good starting points in materials where classical force fields may fall short.

\section{Related Work}

Over the past decade, there has been increasing adoption of machine learning based interatomic potentials for atomistic modeling, with a wide range of architectures proposed in the academic community. Generally, these interatomic potentials have been confined to a small system domain and struggled to generalize across different material classes. MACE \cite{batatia2023macehigherorderequivariant}, the first UIP with significant generalization attributes, has seen widespread usage.

\paragraph{}

Invariance/equivariance with respect to rotation is often considered to be a critical property of such models. Many popular UIPs, such as MACE \cite{batatia2023macehigherorderequivariant}, NEQUIP \cite{nequip} and Equiformer V1/V2 \cite{liao2023equiformerequivariantgraphattention, liao2023equiformerv2} make use of architectures which respect these symmetries by construction. However, respecting symmetries comes at a cost; it requires a unique set of design rules and tensor operations that are challenging to implement efficiently\cite{batatia2022designspacee3equivariantatomcentered}. In contrast, we have opted for a non-equivariant GNS architecture that only deviates in minor ways from the original implementation \cite{SanchezGonzalez2020LearningTS}. Despite the relatively few architectural adjustments,  we have been able to achieve state-of-the-art accuracy and speed.
\paragraph{}

More recently, the benefits of equivariant models have begun to be analyzed and challenged. \cite{langer2024probingeffectsbrokensymmetries} demonstrate that the non-equivariant PET model \cite{pozdnyakov2024smoothexactrotationalsymmetrization} preserves simulation statistics, even in cases where such observables are expected to be affected by broken symmetries. Additionally, in domains outside of atomistic modeling which also deal with point clouds in 3D Euclidean space, unconstrained models are commonplace \cite{qi2017pointnetdeeplearningpoint, zhao2021pointtransformer, wu2020pointconvdeepconvolutionalnetworks}. More recently, non-invariant models have been used for accurately predicting the secondary structure of proteins \cite{alphafold3} and generative models of small molecule conformers \cite{bitterpillconformergeneration}.

\paragraph{}

In parallel with the introduction of modeling techniques which do not respect various Euclidean symmetries, new techniques have been explored for introducing rotational equivariance/invariance at inference time using data augmentation. \cite{gerken2024emergentequivariancedeepensembles} prove that deep ensemble models are equivariant in the limit of infinitely wide network layers when trained with data augmentation, even in off-manifold areas of the data distribution. \cite{kaba2023equivariancelearnedcanonicalizationfunctions} use auxiliary networks to find canonical representations of inputs, leading to exact equivariance. \cite{benton2020learninginvariancesneuralnetworks} discover invariant properties directly from data by learning a joint distribution over model parameters and data augmentation methods. Learned equivariance has been demonstrated to be a function of model size, accuracy and architecture, with \cite{gruver2024liederivativemeasuringlearned} showing that vison transformer models are often more rotationally equivariant than convolutional neural networks, despite the strong translation invariance built into CNNs. This demonstrates that unconstrained models can both learn fundamental invariances from data whilst also being more performant.
\paragraph{}

Molecular dynamics simulations using UIPs are known to be susceptible to instability, causing nonphysical simulation states and simulation collapse. This instability has low correlation with Force and Energy MAE \cite{fu2022forces}, making diagnosing issues with model based simulations difficult. Stability issues are often resolved using active learning to expand the observed potential energy surface \cite{Smith2018LessIM, vandermause2019ontheflyactivelearninginterpretable, zaverkin2023uncertaintybiasedmoleculardynamicslearning, raja2024stabilityawaretrainingmachinelearning}, but this requires expensive rounds of model finetuning, simulations and quantum mechanical calculations. \cite{Stocker_2022} show improved simulation stability with increasing training dataset sizes, which we adopt in combination with noise data augmentations \cite{godwin2022simplegnnregularisation3d} framed as denoising diffusion models \cite{Ho2020DenoisingDP, kingma2023variationaldiffusionmodels}.
\paragraph{}

The majority of prior works in pretraining neural networks for atomistic modeling have focused somewhat narrowly on \emph{either} materials or molecules \cite{zaidi2022pretrainingdenoisingmolecularproperty, godwin2022simplegnnregularisation3d} and adopted model architectures that have yet to exhibit strong data scaling with increases in available training data \cite{shoghi2024moleculesmaterialspretraininglarge, batatia2024foundationmodelatomisticmaterials}. One obvious reason for these shortcomings is the scarcity of compatible DFT data. Our use of generative pretraining alleviates this issue by supporting the combined use of all available DFT datasets, irrespective of level-of-theory. Arguably, we have only scratched the surface of this approach through our use of diffusion modeling on ground-states, and we believe that the use of full trajectory data \cite{liao2024generalizingdenoisingnonequilibriumstructures, noneqdenoisingpretraining} is a promising direction for future work.

\section{Conclusion}

We have presented a universal interatomic potential, Orb, that has reached a unique set of milestones. When released, it set a new state-of-the-art on the competitive Matbench-Discovery benchmark, and continues to be 2-6 times faster than its closest competitors, depending on system size. Moreover, we have found Orb to be stable under simulation of diverse systems, including out-of-distribution small molecules.
\paragraph{}

This paper has explored several strategies for approximating behaviors and properties of both conservative forcefields and models which are equivariant by construction,  whilst exploiting the substantial benefits that come with relaxing these assumptions. Clearly, equivariant models constructed as conservative vector fields are \textit{apriori} better models of atomic forces, but this does not imply that the best learned interatomic potentials will come from this model class. It is this learning step that introduces ambiguity about the correct modeling approach.By removing the mean force vector from force predictions (Eq.\ \ref{eq:mean_force}), we ensure the net force acting on the system is zero - a property of a conservative vector field. By removing torque from predicted forces (Section \ref{section:torque_removal}), we implement an equivalent constraint to an irrotational forcefield, again a property of conservative forcefields. By using data augmentation during training, we produce models which are approximately roto-invariant/equivariant. Importantly, we can achieve these properties (or good approximations to these properties) without complex architectural constraints or expensive gradient computations. 

\paragraph{}
Finally, we have demonstrated Orb's capacity to generalize to new domains with varied chemistries, as well as modifying an existing benchmark, MD-17, as a test of generalization for forcefield models - by introducing a zero-shot evaluation of MD stability and descriptive statistics. We hope that Orb models will find adoption within the computational chemistry community, particularly for simulations that were difficult or impossible to run at the scale required for useful analysis. 

\newpage
\printbibliography

@article{kohn_sham,
  title = {Self-Consistent Equations Including Exchange and Correlation Effects},
  author = {Kohn, W. and Sham, L. J.},
  journal = {Phys. Rev.},
  volume = {140},
  issue = {4A},
  pages = {A1133--A1138},
  numpages = {0},
  year = {1965},
  month = {Nov},
  publisher = {American Physical Society},
  doi = {10.1103/PhysRev.140.A1133},
  url = {https://link.aps.org/doi/10.1103/PhysRev.140.A1133}
}

@article{SanchezGonzalez2020LearningTS,
  title={Learning to Simulate Complex Physics with Graph Networks},
  author={Alvaro Sanchez-Gonzalez and Jonathan Godwin and Tobias Pfaff and Rex Ying and Jure Leskovec and Peter W. Battaglia},
  journal={ArXiv},
  year={2020},
  volume={abs/2002.09405},
  url={https://api.semanticscholar.org/CorpusID:211252550}
}

@article{Velickovic2017GraphAN,
  title={Graph Attention Networks},
  author={Petar Velickovic and Guillem Cucurull and Arantxa Casanova and Adriana Romero and Pietro Lio’ and Yoshua Bengio},
  journal={ArXiv},
  year={2017},
  volume={abs/1710.10903},
  url={https://api.semanticscholar.org/CorpusID:3292002}
}

@article{Ho2020DenoisingDP,
  title={Denoising Diffusion Probabilistic Models},
  author={Jonathan Ho and Ajay Jain and P. Abbeel},
  journal={ArXiv},
  year={2020},
  volume={abs/2006.11239},
  url={https://api.semanticscholar.org/CorpusID:219955663}
}

@inproceedings{Song2019GenerativeMB,
  title={Generative Modeling by Estimating Gradients of the Data Distribution},
  author={Yang Song and Stefano Ermon},
  booktitle={Neural Information Processing Systems},
  year={2019},
  url={https://api.semanticscholar.org/CorpusID:196470871}
}

@article{Ba2016LayerN,
  title={Layer Normalization},
  author={Jimmy Ba and Jamie Ryan Kiros and Geoffrey E. Hinton},
  journal={ArXiv},
  year={2016},
  volume={abs/1607.06450},
  url={https://api.semanticscholar.org/CorpusID:8236317}
}

@article{Bitzek2006StructuralRM,
  title={Structural relaxation made simple.},
  author={Erik Bitzek and Pekka Koskinen and Franz G{\"a}hler and Michael Moseler and Peter Gumbsch},
  journal={Physical review letters},
  year={2006},
  volume={97 17},
  pages={
          170201
        },
  url={https://api.semanticscholar.org/CorpusID:35381339}
}

@article{Grimme2010ACA,
  title={A consistent and accurate ab initio parametrization of density functional dispersion correction (DFT-D) for the 94 elements H-Pu.},
  author={Stefan Grimme and Jens Antony and Stephan Ehrlich and Helge Krieg},
  journal={The Journal of chemical physics},
  year={2010},
  volume={132 15},
  pages={
          154104
        },
  url={https://api.semanticscholar.org/CorpusID:28512828}
}

@article{Perdew1996GeneralizedGA,
  title={Generalized Gradient Approximation Made Simple.},
  author={John P. Perdew and Kieron Burke and Matthias Ernzerhof},
  journal={Physical review letters},
  year={1996},
  volume={77 18},
  pages={
          3865-3868
        },
  url={https://api.semanticscholar.org/CorpusID:6425905}
}

@article{GNOMe,
  title={Scaling deep learning for materials discovery},
  author={Amil Merchant and Simon Batzner and Samuel S. Schoenholz and Muratahan Aykol and Gowoon Cheon and Ekin Dogus Cubuk},
  journal={Nature},
  year={2023},
  volume={624},
  pages={80 - 85},
  url={https://api.semanticscholar.org/CorpusID:265505419}
}

@article{yang2024mattersim,
author = {Yang, Han and Hu, Chenxi and Zhou, Yichi and Liu, Xixian and Shi, Yu and Li, Jielan and Li, Guanzhi and Chen, Zekun and Chen, Shuizhou and Zeni, Claudio and Horton, Matthew and Pinsler, Robert and Fowler, Andrew and Zügner, Daniel and Xie, Tianyidan and Smith, Jake and Sun, Lixin and Wang, Qian and Kong, Lingyu and Liu, Chang and Hao, Hongxia and Lu, Ziheng},
title = {MatterSim: A Deep Learning Atomistic Model Across Elements, Temperatures and Pressures},
howpublished = {arXiv},
year = {2024},
month = {May},
url = {https://www.microsoft.com/en-us/research/publication/mattersim-a-deep-learning-atomistic-model-across-elements-temperatures-and-pressures/},
}

@misc{zeni2024mattergengenerativemodelinorganic,
      title={MatterGen: a generative model for inorganic materials design}, 
      author={Claudio Zeni and Robert Pinsler and Daniel Zügner and Andrew Fowler and Matthew Horton and Xiang Fu and Sasha Shysheya and Jonathan Crabbé and Lixin Sun and Jake Smith and Bichlien Nguyen and Hannes Schulz and Sarah Lewis and Chin-Wei Huang and Ziheng Lu and Yichi Zhou and Han Yang and Hongxia Hao and Jielan Li and Ryota Tomioka and Tian Xie},
      year={2024},
      eprint={2312.03687},
      archivePrefix={arXiv},
      primaryClass={cond-mat.mtrl-sci},
      url={https://arxiv.org/abs/2312.03687}, 
}

@misc{deng2023chgnetpretraineduniversalneural,
      title={CHGNet: Pretrained universal neural network potential for charge-informed atomistic modeling}, 
      author={Bowen Deng and Peichen Zhong and KyuJung Jun and Janosh Riebesell and Kevin Han and Christopher J. Bartel and Gerbrand Ceder},
      year={2023},
      eprint={2302.14231},
      archivePrefix={arXiv},
      primaryClass={cond-mat.mtrl-sci},
      url={https://arxiv.org/abs/2302.14231}, 
}

@article{abinitio_random_struct_search,
   title={Ab initio random structure searching},
   volume={23},
   ISSN={1361-648X},
   url={http://dx.doi.org/10.1088/0953-8984/23/5/053201},
   DOI={10.1088/0953-8984/23/5/053201},
   number={5},
   journal={Journal of Physics: Condensed Matter},
   publisher={IOP Publishing},
   author={Pickard, Chris J and Needs, R J},
   year={2011},
   month=jan, pages={053201} }

@misc{riebesell2024matbenchdiscoveryframework,
      title={Matbench Discovery -- A framework to evaluate machine learning crystal stability predictions}, 
      author={Janosh Riebesell and Rhys E. A. Goodall and Philipp Benner and Yuan Chiang and Bowen Deng and Alpha A. Lee and Anubhav Jain and Kristin A. Persson},
      year={2024},
      eprint={2308.14920},
      archivePrefix={arXiv},
      primaryClass={cond-mat.mtrl-sci},
      url={https://arxiv.org/abs/2308.14920}, 
}

@misc{kingma2017adammethodstochasticoptimization,
      title={Adam: A Method for Stochastic Optimization}, 
      author={Diederik P. Kingma and Jimmy Ba},
      year={2017},
      eprint={1412.6980},
      archivePrefix={arXiv},
      primaryClass={cs.LG},
      url={https://arxiv.org/abs/1412.6980}, 
}

@misc{paszke2019pytorchimperativestylehighperformance,
      title={PyTorch: An Imperative Style, High-Performance Deep Learning Library}, 
      author={Adam Paszke and Sam Gross and Francisco Massa and Adam Lerer and James Bradbury and Gregory Chanan and Trevor Killeen and Zeming Lin and Natalia Gimelshein and Luca Antiga and Alban Desmaison and Andreas Köpf and Edward Yang and Zach DeVito and Martin Raison and Alykhan Tejani and Sasank Chilamkurthy and Benoit Steiner and Lu Fang and Junjie Bai and Soumith Chintala},
      year={2019},
      eprint={1912.01703},
      archivePrefix={arXiv},
      primaryClass={cs.LG},
      url={https://arxiv.org/abs/1912.01703}, 
}

@misc{kingma2023variationaldiffusionmodels,
      title={Variational Diffusion Models}, 
      author={Diederik P. Kingma and Tim Salimans and Ben Poole and Jonathan Ho},
      year={2023},
      eprint={2107.00630},
      archivePrefix={arXiv},
      primaryClass={cs.LG},
      url={https://arxiv.org/abs/2107.00630}, 
}

@misc{langer2024probingeffectsbrokensymmetries,
      title={Probing the effects of broken symmetries in machine learning}, 
      author={Marcel F. Langer and Sergey N. Pozdnyakov and Michele Ceriotti},
      year={2024},
      eprint={2406.17747},
      archivePrefix={arXiv},
      primaryClass={physics.chem-ph},
      url={https://arxiv.org/abs/2406.17747}, 
}

@misc{pozdnyakov2024smoothexactrotationalsymmetrization,
      title={Smooth, exact rotational symmetrization for deep learning on point clouds}, 
      author={Sergey N. Pozdnyakov and Michele Ceriotti},
      year={2024},
      eprint={2305.19302},
      archivePrefix={arXiv},
      primaryClass={cs.CV},
      url={https://arxiv.org/abs/2305.19302}, 
}

@misc{batatia2022designspacee3equivariantatomcentered,
      title={The Design Space of E(3)-Equivariant Atom-Centered Interatomic Potentials}, 
      author={Ilyes Batatia and Simon Batzner and Dávid Péter Kovács and Albert Musaelian and Gregor N. C. Simm and Ralf Drautz and Christoph Ortner and Boris Kozinsky and Gábor Csányi},
      year={2022},
      eprint={2205.06643},
      archivePrefix={arXiv},
      primaryClass={stat.ML},
      url={https://arxiv.org/abs/2205.06643}, 
}

@misc{batatia2023macehigherorderequivariant,
      title={MACE: Higher Order Equivariant Message Passing Neural Networks for Fast and Accurate Force Fields}, 
      author={Ilyes Batatia and Dávid Péter Kovács and Gregor N. C. Simm and Christoph Ortner and Gábor Csányi},
      year={2023},
      eprint={2206.07697},
      archivePrefix={arXiv},
      primaryClass={stat.ML},
      url={https://arxiv.org/abs/2206.07697}, 
}

@article{nequip,
   title={E(3)-equivariant graph neural networks for data-efficient and accurate interatomic potentials},
   volume={13},
   ISSN={2041-1723},
   url={http://dx.doi.org/10.1038/s41467-022-29939-5},
   DOI={10.1038/s41467-022-29939-5},
   number={1},
   journal={Nature Communications},
   publisher={Springer Science and Business Media LLC},
   author={Batzner, Simon and Musaelian, Albert and Sun, Lixin and Geiger, Mario and Mailoa, Jonathan P. and Kornbluth, Mordechai and Molinari, Nicola and Smidt, Tess E. and Kozinsky, Boris},
   year={2022},
   month=may }

@misc{liao2023equiformerequivariantgraphattention,
      title={Equiformer: Equivariant Graph Attention Transformer for 3D Atomistic Graphs}, 
      author={Yi-Lun Liao and Tess Smidt},
      year={2023},
      eprint={2206.11990},
      archivePrefix={arXiv},
      primaryClass={cs.LG},
      url={https://arxiv.org/abs/2206.11990}, 
}

@misc{qi2017pointnetdeeplearningpoint,
      title={PointNet: Deep Learning on Point Sets for 3D Classification and Segmentation}, 
      author={Charles R. Qi and Hao Su and Kaichun Mo and Leonidas J. Guibas},
      year={2017},
      eprint={1612.00593},
      archivePrefix={arXiv},
      primaryClass={cs.CV},
      url={https://arxiv.org/abs/1612.00593}, 
}

@misc{zhao2021pointtransformer,
      title={Point Transformer}, 
      author={Hengshuang Zhao and Li Jiang and Jiaya Jia and Philip Torr and Vladlen Koltun},
      year={2021},
      eprint={2012.09164},
      archivePrefix={arXiv},
      primaryClass={cs.CV},
      url={https://arxiv.org/abs/2012.09164}, 
}

@misc{wu2020pointconvdeepconvolutionalnetworks,
      title={PointConv: Deep Convolutional Networks on 3D Point Clouds}, 
      author={Wenxuan Wu and Zhongang Qi and Li Fuxin},
      year={2020},
      eprint={1811.07246},
      archivePrefix={arXiv},
      primaryClass={cs.CV},
      url={https://arxiv.org/abs/1811.07246}, 
}

@article{alphafold3,
  title={Accurate structure prediction of biomolecular interactions with AlphaFold 3},
  author={Josh Abramson and Jonas Adler and Jack Dunger and Richard Evans and Tim Green and Alexander Pritzel and Olaf Ronneberger and Lindsay Willmore and Andrew J Ballard and Joshua Bambrick and Sebastian W Bodenstein and David A Evans and Chia-Chun Hung and Michael O’Neill and David Reiman and Kathryn Tunyasuvunakool and Zachary Wu and Akvilė Žemgulytė and Eirini Arvaniti and Charles Beattie and Ottavia Bertolli and Alex Bridgland and Alexey Cherepanov and Miles Congreve and Alexander Imani Cowen-Rivers and Andrew Cowie and Michael Figurnov and Fabian B Fuchs and Hannah Gladman and Rishub Jain and Yousuf A. Khan and Caroline M R Low and Kuba Perlin and Anna Potapenko and Pascal Savy and Sukhdeep Singh and Adrian Stecula and Ashok Thillaisundaram and Catherine Tong and Sergei Yakneen and Ellen D. Zhong and Michal Zielinski and Augustin Ž{\'i}dek and Vic-613 tor Bapst and Pushmeet Kohli and Max Jaderberg and Demis Hassabis and John M. Jumper},
  journal={Nature},
  year={2024},
  volume={630},
  pages={493 - 500},
  url={https://api.semanticscholar.org/CorpusID:269633210}
}

@misc{bitterpillconformergeneration,
      title={Swallowing the Bitter Pill: Simplified Scalable Conformer Generation}, 
      author={Yuyang Wang and Ahmed A. Elhag and Navdeep Jaitly and Joshua M. Susskind and Miguel Angel Bautista},
      year={2024},
      eprint={2311.17932},
      archivePrefix={arXiv},
      primaryClass={physics.chem-ph},
      url={https://arxiv.org/abs/2311.17932}, 
}

@misc{gerken2024emergentequivariancedeepensembles,
      title={Emergent Equivariance in Deep Ensembles}, 
      author={Jan E. Gerken and Pan Kessel},
      year={2024},
      eprint={2403.03103},
      archivePrefix={arXiv},
      primaryClass={cs.LG},
      url={https://arxiv.org/abs/2403.03103}, 
}

@misc{kaba2023equivariancelearnedcanonicalizationfunctions,
      title={Equivariance with Learned Canonicalization Functions}, 
      author={Sékou-Oumar Kaba and Arnab Kumar Mondal and Yan Zhang and Yoshua Bengio and Siamak Ravanbakhsh},
      year={2023},
      eprint={2211.06489},
      archivePrefix={arXiv},
      primaryClass={cs.LG},
      url={https://arxiv.org/abs/2211.06489}, 
}

@misc{benton2020learninginvariancesneuralnetworks,
      title={Learning Invariances in Neural Networks}, 
      author={Gregory Benton and Marc Finzi and Pavel Izmailov and Andrew Gordon Wilson},
      year={2020},
      eprint={2010.11882},
      archivePrefix={arXiv},
      primaryClass={cs.LG},
      url={https://arxiv.org/abs/2010.11882}, 
}

@article{Smith2018LessIM,
  title={Less is more: sampling chemical space with active learning},
  author={Justin S. Smith and Benjamin Tyler Nebgen and Nicholas Lubbers and Olexandr Isayev and Adrian E. Roitberg},
  journal={The Journal of chemical physics},
  year={2018},
  volume={148 24},
  pages={
          241733
        },
  url={https://api.semanticscholar.org/CorpusID:4682180}
}

@misc{vandermause2019ontheflyactivelearninginterpretable,
      title={On-the-Fly Active Learning of Interpretable Bayesian Force Fields for Atomistic Rare Events}, 
      author={Jonathan Vandermause and Steven B. Torrisi and Simon Batzner and Yu Xie and Lixin Sun and Alexie M. Kolpak and Boris Kozinsky},
      year={2019},
      eprint={1904.02042},
      archivePrefix={arXiv},
      primaryClass={physics.comp-ph},
      url={https://arxiv.org/abs/1904.02042}, 
}

@misc{zaverkin2023uncertaintybiasedmoleculardynamicslearning,
      title={Uncertainty-biased molecular dynamics for learning uniformly accurate interatomic potentials}, 
      author={Viktor Zaverkin and David Holzmüller and Henrik Christiansen and Federico Errica and Francesco Alesiani and Makoto Takamoto and Mathias Niepert and Johannes Kästner},
      year={2023},
      eprint={2312.01416},
      archivePrefix={arXiv},
      primaryClass={physics.comp-ph},
      url={https://arxiv.org/abs/2312.01416}, 
}

@misc{raja2024stabilityawaretrainingmachinelearning,
      title={Stability-Aware Training of Machine Learning Force Fields with Differentiable Boltzmann Estimators}, 
      author={Sanjeev Raja and Ishan Amin and Fabian Pedregosa and Aditi S. Krishnapriyan},
      year={2024},
      eprint={2402.13984},
      archivePrefix={arXiv},
      primaryClass={cs.LG},
      url={https://arxiv.org/abs/2402.13984}, 
}

@article{Stocker_2022,
doi = {10.1088/2632-2153/ac9955},
url = {https://dx.doi.org/10.1088/2632-2153/ac9955},
year = {2022},
month = {nov},
publisher = {IOP Publishing},
volume = {3},
number = {4},
pages = {045010},
author = {Sina Stocker and Johannes Gasteiger and Florian Becker and Stephan Günnemann and Johannes T Margraf},
title = {How robust are modern graph neural network potentials in long and hot molecular dynamics simulations?},
journal = {Machine Learning: Science and Technology},
}

@misc{godwin2022simplegnnregularisation3d,
      title={Simple GNN Regularisation for 3D Molecular Property Prediction \& Beyond}, 
      author={Jonathan Godwin and Michael Schaarschmidt and Alexander Gaunt and Alvaro Sanchez-Gonzalez and Yulia Rubanova and Petar Veličković and James Kirkpatrick and Peter Battaglia},
      year={2022},
      eprint={2106.07971},
      archivePrefix={arXiv},
      primaryClass={cs.LG},
      url={https://arxiv.org/abs/2106.07971}, 
}

@misc{zaidi2022pretrainingdenoisingmolecularproperty,
      title={Pre-training via Denoising for Molecular Property Prediction}, 
      author={Sheheryar Zaidi and Michael Schaarschmidt and James Martens and Hyunjik Kim and Yee Whye Teh and Alvaro Sanchez-Gonzalez and Peter Battaglia and Razvan Pascanu and Jonathan Godwin},
      year={2022},
      eprint={2206.00133},
      archivePrefix={arXiv},
      primaryClass={cs.LG},
      url={https://arxiv.org/abs/2206.00133}, 
}

@misc{shoghi2024moleculesmaterialspretraininglarge,
      title={From Molecules to Materials: Pre-training Large Generalizable Models for Atomic Property Prediction}, 
      author={Nima Shoghi and Adeesh Kolluru and John R. Kitchin and Zachary W. Ulissi and C. Lawrence Zitnick and Brandon M. Wood},
      year={2024},
      eprint={2310.16802},
      archivePrefix={arXiv},
      primaryClass={cs.LG},
      url={https://arxiv.org/abs/2310.16802}, 
}

@misc{barrosoluque2024openmaterials2024omat24,
      title={Open Materials 2024 (OMat24) Inorganic Materials Dataset and Models}, 
      author={Luis Barroso-Luque and Muhammed Shuaibi and Xiang Fu and Brandon M. Wood and Misko Dzamba and Meng Gao and Ammar Rizvi and C. Lawrence Zitnick and Zachary W. Ulissi},
      year={2024},
      eprint={2410.12771},
      archivePrefix={arXiv},
      primaryClass={cond-mat.mtrl-sci},
      url={https://arxiv.org/abs/2410.12771}, 
}

@misc{batatia2024foundationmodelatomisticmaterials,
      title={A foundation model for atomistic materials chemistry}, 
      author={Ilyes Batatia and Philipp Benner and Yuan Chiang and Alin M. Elena and Dávid P. Kovács and Janosh Riebesell and Xavier R. Advincula and Mark Asta and Matthew Avaylon and William J. Baldwin and Fabian Berger and Noam Bernstein and Arghya Bhowmik and Samuel M. Blau and Vlad Cărare and James P. Darby and Sandip De and Flaviano Della Pia and Volker L. Deringer and Rokas Elijošius and Zakariya El-Machachi and Fabio Falcioni and Edvin Fako and Andrea C. Ferrari and Annalena Genreith-Schriever and Janine George and Rhys E. A. Goodall and Clare P. Grey and Petr Grigorev and Shuang Han and Will Handley and Hendrik H. Heenen and Kersti Hermansson and Christian Holm and Jad Jaafar and Stephan Hofmann and Konstantin S. Jakob and Hyunwook Jung and Venkat Kapil and Aaron D. Kaplan and Nima Karimitari and James R. Kermode and Namu Kroupa and Jolla Kullgren and Matthew C. Kuner and Domantas Kuryla and Guoda Liepuoniute and Johannes T. Margraf and Ioan-Bogdan Magdău and Angelos Michaelides and J. Harry Moore and Aakash A. Naik and Samuel P. Niblett and Sam Walton Norwood and Niamh O'Neill and Christoph Ortner and Kristin A. Persson and Karsten Reuter and Andrew S. Rosen and Lars L. Schaaf and Christoph Schran and Benjamin X. Shi and Eric Sivonxay and Tamás K. Stenczel and Viktor Svahn and Christopher Sutton and Thomas D. Swinburne and Jules Tilly and Cas van der Oord and Eszter Varga-Umbrich and Tejs Vegge and Martin Vondrák and Yangshuai Wang and William C. Witt and Fabian Zills and Gábor Csányi},
      year={2024},
      eprint={2401.00096},
      archivePrefix={arXiv},
      primaryClass={physics.chem-ph},
      url={https://arxiv.org/abs/2401.00096}, 
}

@misc{liao2024generalizingdenoisingnonequilibriumstructures,
      title={Generalizing Denoising to Non-Equilibrium Structures Improves Equivariant Force Fields}, 
      author={Yi-Lun Liao and Tess Smidt and Muhammed Shuaibi and Abhishek Das},
      year={2024},
      eprint={2403.09549},
      archivePrefix={arXiv},
      primaryClass={cs.LG},
      url={https://arxiv.org/abs/2403.09549}, 
}

@article{noneqdenoisingpretraining,
   title={Denoise Pretraining on Nonequilibrium Molecules for Accurate and Transferable Neural Potentials},
   volume={19},
   ISSN={1549-9626},
   url={http://dx.doi.org/10.1021/acs.jctc.3c00289},
   DOI={10.1021/acs.jctc.3c00289},
   number={15},
   journal={Journal of Chemical Theory and Computation},
   publisher={American Chemical Society (ACS)},
   author={Wang, Yuyang and Xu, Changwen and Li, Zijie and Barati Farimani, Amir},
   year={2023},
   month=jun, pages={5077–5087} }

@misc{vaswani2023attentionneed,
      title={Attention Is All You Need}, 
      author={Ashish Vaswani and Noam Shazeer and Niki Parmar and Jakob Uszkoreit and Llion Jones and Aidan N. Gomez and Lukasz Kaiser and Illia Polosukhin},
      year={2023},
      eprint={1706.03762},
      archivePrefix={arXiv},
      primaryClass={cs.CL},
      url={https://arxiv.org/abs/1706.03762}, 
}

@misc{li2018deeperinsightsgraphconvolutional,
      title={Deeper Insights into Graph Convolutional Networks for Semi-Supervised Learning}, 
      author={Qimai Li and Zhichao Han and Xiao-Ming Wu},
      year={2018},
      eprint={1801.07606},
      archivePrefix={arXiv},
      primaryClass={cs.LG},
      url={https://arxiv.org/abs/1801.07606}, 
}

@ARTICLE{gnns,
  author={Scarselli, Franco and Gori, Marco and Tsoi, Ah Chung and Hagenbuchner, Markus and Monfardini, Gabriele},
  journal={IEEE Transactions on Neural Networks}, 
  title={The Graph Neural Network Model}, 
  year={2009},
  volume={20},
  number={1},
  pages={61-80},
  doi={10.1109/TNN.2008.2005605}}

@misc{gruver2024liederivativemeasuringlearned,
      title={The Lie Derivative for Measuring Learned Equivariance}, 
      author={Nate Gruver and Marc Finzi and Micah Goldblum and Andrew Gordon Wilson},
      year={2024},
      eprint={2210.02984},
      archivePrefix={arXiv},
      primaryClass={cs.LG},
      url={https://arxiv.org/abs/2210.02984}, 
}

@article{OQMD,
  journal={NPJ Computational Materials},
  title={The Open Quantum Materials Database (OQMD): assessing the accuracy of DFT formation energies},
  author={Scott Kirklin and James Edward Saal and Bryce Meredig and Alexander Thompson and Jeff W. Doak and Muratahan Aykol and Stephan Ruhl and Christopher M. Wolverton},
  year={2015},
  url={https://api.semanticscholar.org/CorpusID:125046626}
}

@article{alexandria,
author = {Schmidt, Jonathan and Hoffmann, Noah and Wang, Hai-Chen and Borlido, Pedro and Carriço, Pedro J. M. A. and Cerqueira, Tiago F. T. and Botti, Silvana and Marques, Miguel A. L.},
title = {Machine-Learning-Assisted Determination of the Global Zero-Temperature Phase Diagram of Materials},
journal = {Advanced Materials},
volume = {35},
number = {22},
pages = {2210788},
doi = {https://doi.org/10.1002/adma.202210788},
url = {https://onlinelibrary.wiley.com/doi/abs/10.1002/adma.202210788},
eprint = {https://onlinelibrary.wiley.com/doi/pdf/10.1002/adma.202210788},
year = {2023}
}

@inproceedings{sohl2015deep,
  title={Deep unsupervised learning using nonequilibrium thermodynamics},
  author={Sohl-Dickstein, Jascha and Weiss, Eric and Maheswaranathan, Niru and Ganguli, Surya},
  booktitle={International conference on machine learning},
  pages={2256--2265},
  year={2015},
  organization={PMLR}
}

@article{vincent2011connection,
  title={A connection between score matching and denoising autoencoders},
  author={Vincent, Pascal},
  journal={Neural computation},
  volume={23},
  number={7},
  pages={1661--1674},
  year={2011},
  publisher={MIT Press}
}

@article{fu2022forces,
  title={Forces are not enough: Benchmark and critical evaluation for machine learning force fields with molecular simulations},
  author={Fu, Xiang and Wu, Zhenghao and Wang, Wujie and Xie, Tian and Keten, Sinan and Gomez-Bombarelli, Rafael and Jaakkola, Tommi},
  journal={arXiv preprint arXiv:2210.07237},
  year={2022}
}

@article{hyvarinen2005estimation,
  title={Estimation of non-normalized statistical models by score matching.},
  author={Hyv{\"a}rinen, Aapo and Dayan, Peter},
  journal={Journal of Machine Learning Research},
  volume={6},
  number={4},
  year={2005}
}

@article{queen2014comprehensive,
  title={Comprehensive study of carbon dioxide adsorption in the metal--organic frameworks M 2 (dobdc)(M= Mg, Mn, Fe, Co, Ni, Cu, Zn)},
  author={Queen, Wendy L and Hudson, Matthew R and Bloch, Eric D and Mason, Jarad A and Gonzalez, Miguel I and Lee, Jason S and Gygi, David and Howe, Joshua D and Lee, Kyuho and Darwish, Tamim A and others},
  journal={Chemical Science},
  volume={5},
  number={12},
  pages={4569--4581},
  year={2014},
  publisher={Royal Society of Chemistry}
}

@article{caskey2008dramatic,
  title={Dramatic tuning of carbon dioxide uptake via metal substitution in a coordination polymer with cylindrical pores},
  author={Caskey, Stephen R and Wong-Foy, Antek G and Matzger, Adam J},
  journal={Journal of the American Chemical Society},
  volume={130},
  number={33},
  pages={10870--10871},
  year={2008},
  publisher={ACS Publications}
}

@article{widom1963some,
  title={Some topics in the theory of fluids},
  author={Widom, Ben},
  journal={The Journal of Chemical Physics},
  volume={39},
  number={11},
  pages={2808--2812},
  year={1963},
  publisher={American Institute of Physics}
}

@article{vandenbrande2018ab,
  title={Ab initio evaluation of Henry coefficients using importance sampling},
  author={Vandenbrande, Steven and Waroquier, Michel and Van Speybroeck, Veronique and Verstraelen, Toon},
  journal={Journal of chemical theory and computation},
  volume={14},
  number={12},
  pages={6359--6369},
  year={2018},
  publisher={ACS Publications}
}

@article{liao2023equiformerv2,
  title={Equiformerv2: Improved equivariant transformer for scaling to higher-degree representations},
  author={Liao, Yi-Lun and Wood, Brandon and Das, Abhishek and Smidt, Tess},
  journal={arXiv preprint arXiv:2306.12059},
  year={2023}
}

@article{goeminne2023dft,
  title={DFT-Quality adsorption simulations in metal--organic frameworks enabled by machine learning Potentials},
  author={Goeminne, Ruben and Vanduyfhuys, Louis and Van Speybroeck, Veronique and Verstraelen, Toon},
  journal={Journal of Chemical Theory and Computation},
  volume={19},
  number={18},
  pages={6313--6325},
  year={2023},
  publisher={ACS Publications}
}

@article{yang2011methyl,
  title={Methyl modified MOF-5: a water stable hydrogen storage material},
  author={Yang, Jie and Grzech, Anna and Mulder, Fokko M and Dingemans, Theo J},
  journal={Chemical communications},
  volume={47},
  number={18},
  pages={5244--5246},
  year={2011},
  publisher={Royal Society of Chemistry}
}

@article{lock2010elucidating,
  title={Elucidating negative thermal expansion in MOF-5},
  author={Lock, Nina and Wu, Yue and Christensen, Mogens and Cameron, Lisa J and Peterson, Vanessa K and Bridgeman, Adam J and Kepert, Cameron J and Iversen, Bo B},
  journal={The Journal of Physical Chemistry C},
  volume={114},
  number={39},
  pages={16181--16186},
  year={2010},
  publisher={ACS Publications}
}

\section{Appendix}

\subsection{Code Availability}

Model weights and code are available under an Apache 2.0 License on Github at \url{https://github.com/orbital-materials/orb-models}.

\subsection{Hyperparameters}

Models are implemented in Pytorch \cite{paszke2019pytorchimperativestylehighperformance} and trained using ADAM \cite{kingma2017adammethodstochasticoptimization} with a learning rate of $3e^-4$ and $\beta_1 = 0.9$, $\beta_2  = 0.999$ on 8 NVIDIA A100 GPUs. Diffusion pretraining uses the \texttt{CosineAnnealingLR} learning rate schedule and is trained to a max of 200 epochs or until convergence, whichever is sooner. Finetuning uses the \texttt{CosineAnnealing} learning rate schedule with a max learning rate of $3e^-3$ and a warmup of 5\% of the total number of gradient steps. Models are finetuned on a single A100 GPU. All models use an exponentially weighted model average (EMA) with a decay rate of 0.999. Diffusion models use a low discrepancy noise sampler in order to ensure a uniform distribution of noise timesteps per batch \cite{kingma2023variationaldiffusionmodels}.

All models use random rotation augmentations during training. Atomic graph edges are featurized using the 20 nearest neighbors of a max radius of 10. Systems with unit cells use a Minimum Image Convention aware implementation of nearest neighbors. We use a dynamic batch size based on a maximum number of nodes, edges and graphs, resulting in an average per GPU batch size of 240 systems.

\subsection{Net Force and Net Torque Removal}
\label{section:torque_removal}
As explained in the main text, our model is non-conservative and thus directly predicts force vectors rather than computing gradients of an energy function. This is not uncommon in the literature, and can work very well as measured by MAE on a test set, or when optimising structures as in the Matbench-Discovery benchmark.

However, we have found that na{\"i}vely predicting unconstrained forces can be very unstable in Molecular Dynamics simulations. Systems can undergo strong net translational or rotational forces that are not possible in real physical systems in a static equilibrium. 

We have found two modifications to the force predictions to be effective for resolving this. The first modification is a mean-subtraction:
\begin{equation}
\mathbf{f}_i^{\text{pred}} = \mathbf{f}_i - \frac{1}{N} \sum_{i=1}^{N} \mathbf{f}_i. \label{eq:mean_force}    
\end{equation}

where \(\mathbf{f}_i \in \mathbb{R}^{3}\) is the direct output of the model's \(MLP_\text{forces}\) head. Mean-centering in this fashion ensures no net force acting on the system.

The second modification is to remove net torque in \emph{non-pbc} systems. Again, this is a property dictated by fundamental physics for isolated molecules in static equilibrium; there should not be a `net rotation' of the system. Our approach is to formulate the problem as a constrained optimization problem that can be solved efficiently with Lagrange multipliers.

The solution is inexpensive to compute and can be applied \emph{purely at inference time} to substantially enhance the quality of MD simulations on non-pbc systems. \\

\noindent \textbf{Problem Statement} \\[0.5em]
Given a set of $N$ atoms with positions $\mathbf{R}_i \in \mathbb{R}^3$, for $i = 1, 2, \dots, N$ and predicted forces $\mathbf{f}_i \in \mathbb{R}^3$, our goal is to find an additive adjustment $\delta \mathbf{f}_i \in \mathbb{R}^3$ to the forces with minimal L2 norm subject to the constraints:
\begin{enumerate}
    \item The average net adjustment is zero:
    \[
    \sum_{i=1}^N \delta \mathbf{f}_i = \mathbf{0}  \hspace{10mm} (\text{Constraint $\mathbf{C}_1$})
    \]
    \item The net torque of the adjustment cancels the net torque of the original forces:
    \[
    \sum_{i=1}^N \mathbf{r}_i \times \delta \mathbf{f}_i = -\boldsymbol{\tau}_{\text{total}} \hspace{10mm} (\text{Constraint $\mathbf{C}_2$})
    \]
    where:
    \begin{itemize}
        \item $\mathbf{r}_i = \mathbf{R}_i - \mathbf{R}_{\text{COM}}$ is the position of atom $i$ relative to the center of mass.
        \item $\mathbf{R}_{\text{COM}} = \frac{1}{N} \sum_{i=1}^N \mathbf{R}_i$ is the center of mass of the system.
        \item $\boldsymbol{\tau}_{\text{total}} = \sum_{i=1}^N \mathbf{r}_i \times \mathbf{f}_i$ is the net torque of the predicted forces.
    \end{itemize}
\end{enumerate}

\noindent \textbf{Solving the Constrained Minimization Problem} \\[0.5em]
To solve this constrained optimization problem, we introduce Lagrange multipliers:
\begin{itemize}
    \item $\boldsymbol{\lambda} \in \mathbb{R}^3$ corresponding to the net force constraint $\mathbf{C}_1$.
    \item $\boldsymbol{\mu} \in \mathbb{R}^3$ corresponding to the net torque constraint $\mathbf{C}_2$.
\end{itemize}
yielding the Lagrangian:
\[
\mathcal{L} = \frac{1}{2}\sum_{i=1}^N \|\delta \mathbf{f}_i\|^2 - \boldsymbol{\lambda} \cdot \left( \sum_{i=1}^N \delta \mathbf{f}_i \right) - \boldsymbol{\mu} \cdot \left( \sum_{i=1}^N \mathbf{r}_i \times \delta \mathbf{f}_i + \boldsymbol{\tau}_{\text{total}} \right)
\] \\
Differentiating with respect to each $\delta \mathbf{f}_i$ and solving for when the derivatives are zero gives:
\[
\delta \mathbf{f}_i =  \boldsymbol{\lambda} + \boldsymbol{\mu} \times \mathbf{r}_i .
\]
Thus the only task that remains is to use the constraints to solve for $\boldsymbol{\lambda}$ and $\boldsymbol{\mu}$. \\

\noindent \textbf{Solving for $\boldsymbol{\lambda}$ using Constraint $\mathbf{C}_1$:} \\[0.5em]
\[
\sum_{i=1}^N \delta \mathbf{f}_i = \sum_{i=1}^N \left( \boldsymbol{\lambda} + \boldsymbol{\mu} \times \mathbf{r}_i \right) = N \boldsymbol{\lambda} + \boldsymbol{\mu} \times \left( \sum_{i=1}^N \mathbf{r}_i \right) = \mathbf{0}
\]
But since:
\[
\sum_{i=1}^N \mathbf{r}_i = \sum_{i=1}^N \left( \mathbf{R}_i - \mathbf{R}_{\text{COM}} \right) = N \mathbf{R}_{\text{COM}} - N \mathbf{R}_{\text{COM}} = \mathbf{0}
\]
Thus, the second term vanishes, and we have
\[
N \boldsymbol{\lambda} = \mathbf{0} \implies \boldsymbol{\lambda} = \mathbf{0}
\]
\noindent \textbf{Solving for $\boldsymbol{\mu}$ using Constraint $\mathbf{C}_2$:} \\[0.5em]
\[
\sum_{i=1}^N \mathbf{r}_i \times \delta \mathbf{f}_i = \sum_{i=1}^N \mathbf{r}_i \times \left( \boldsymbol{\mu} \times \mathbf{r}_i \right) = -\boldsymbol{\tau}_{\text{total}}
\]
Applying a standard triple vector product identity, we get
\[
\sum_{i=1}^N \mathbf{r}_i \times \left( \boldsymbol{\mu} \times \mathbf{r}_i \right) = \boldsymbol{\mu} \sum_{i=1}^N \left( \mathbf{r}_i \cdot \mathbf{r}_i \right) - \sum_{i=1}^N \mathbf{r}_i \left( \mathbf{r}_i \cdot \boldsymbol{\mu} \right) = -\boldsymbol{\tau}_{\text{total}}
\]
To simplify this further, let us introduce the notation
\[
    s := \sum_{i=1}^N \|\mathbf{r}_i\|^2, \hspace{10mm} S := \sum_{i=1}^N \mathbf{r}_i \mathbf{r}_i^T
\]
Then, the torque constraint becomes
\begin{flalign*}
&\hspace{10em} \boldsymbol{\mu} s - S \boldsymbol{\mu} = -\boldsymbol{\tau}_{\text{total}} & \\
\Longrightarrow &\hspace{10em} \left( S - s \mathbf{I} \right) \boldsymbol{\mu} = -\boldsymbol{\tau}_{\text{total}} & \\
\Longrightarrow &\hspace{10em} \boldsymbol{\mu} = -(S - s \mathbf{I})^{-1} \boldsymbol{\tau}_{\text{total}} &
\end{flalign*}
Where $\mathbf{I}$ is the $3\times3$ identity matrix. We have assumed here that $S - s \mathbf{I}$ is invertible; if it is instead singular or nearly singular, we use the Moore-Penrose pseudo-inverse. \\

\noindent \textbf{Final expression for adjusted forces} \\[0.5em]
Putting everything together, the final adjusted forces with net zero torque are:
\[
\mathbf{f}_i^{\text{adjusted}} = \mathbf{f}_i + \delta \mathbf{f}_i = \mathbf{f}_i -(S - s \mathbf{I})^{-1} \boldsymbol{\tau}_{\text{total}} \times \mathbf{r}_i
\]

\subsection{Molecular Dynamics}

For evaluating the thermal stability of MOF-5, MD simulations were performed at a constant volume using the NPT ensemble with no external stress applied.
Specifically, we use ASE NPT implementation (version $3.23.0$) 
of Nose-Hoover and Parrinello-Rahman dynamics along with the Orb-D3 model.
Simulations were conducted over a temperature range from 300 K to 1,000 K. To investigate the stability at each temperature step, velocities were initialized from a Boltzmann distribution at each step using a 1 fs timestep. States were sampled every 100 steps along the trajectory.

\subsection{Applications}

Using Widom insertion,\cite{widom1963some} we use a sum approximation to the continuous ensemble average energy to calculate the adsorption energy (Equation \ref{eq:ads_energy}).

\begin{equation}
    \Delta H_{\text{ads}} = \frac{\sum_{i} \Delta E_i \exp\left(-\frac{\Delta E_i}{RT}\right)}{\sum_{i} \exp\left(-\frac{\Delta E_i}{RT}\right)} - RT
    \label{eq:ads_energy}
\end{equation}

Here, $ \Delta E_i $ is the energy change associated with inserting a CO2 molecule into the system at position \textit{i}, \textit{R} is the ideal gas constant, and \textit{T} is the temperature.
By sampling a large number of random insertions, the value converges to the adsorption energy at infinite dilution. 
For computational efficiency and to avoid erroneous predictions due to close range interactions, we use an exclusion sphere of the covalent radius plus 0.2 Å, which allows for placement of the  $\text{CO}_2$ within close proximity of the metal center. 
For both MACE and ORB, we include D3 corrections and each heat of adsorption value was computed following 50,000 trial insertions. We use the structure of Mg-MOF-74 from Goeminne \textit{et al.} optimized at the PBE+D3 level of theory for performing insertions.\cite{goeminne2023dft} 
To compute the free energy surface in Figure \ref{fig:mof-adsorption}, we use

\begin{equation}
    F = -k_B T \ln \left( \sum_{i} e^{-\Delta E_i} \right)
\end{equation}

where \textit{F} is the free energy associated with a rotational state, $k_B$ is the Boltzmann factor, \textit{T} is the temperature, $\beta$ is the inverse temperature, and  $ \Delta E_i $ is the energy of the \textit{i}-th rotational state. Free energies were computed by discretizing the volume into a 20 x 20 grid and using linear interpolation between adjacent points. All calculations for free energy and heat of adsorption were conducted at 298 K.

\subsection{OMAT 24 Dataset and Models}
\label{appendix:omat}

The OMAT 24 dataset \cite{barrosoluque2024openmaterials2024omat24} were released 16th Oct 2024, after the release of Orb, but before the release of this technical report. EquiformerV2 \cite{liao2023equiformerv2} models trained on this dataset surpass the performance of Orb-v2 on the Matbench Discovery benchmark. We believe the OMAT 24 dataset and methods are complimentary to Orb; comparable model sizes (eqV2-S) with similar datasets (MPTraj only) and training methods which do not include DeNS as as fine-tuning time data augmentation method \cite{noneqdenoisingpretraining} perform similarly to Orb-v2, whilst using 8x more GPU resources to train. DeNS trajectory pretraining and the use of OMAT are both directly applicable to Orb models; this is a direction we will explore in future releases.

\subsection{Homo-Nuclear Diatomics}

\begin{figure}
    \makebox[\textwidth][c]{
        \includegraphics[width=1.2\textwidth]{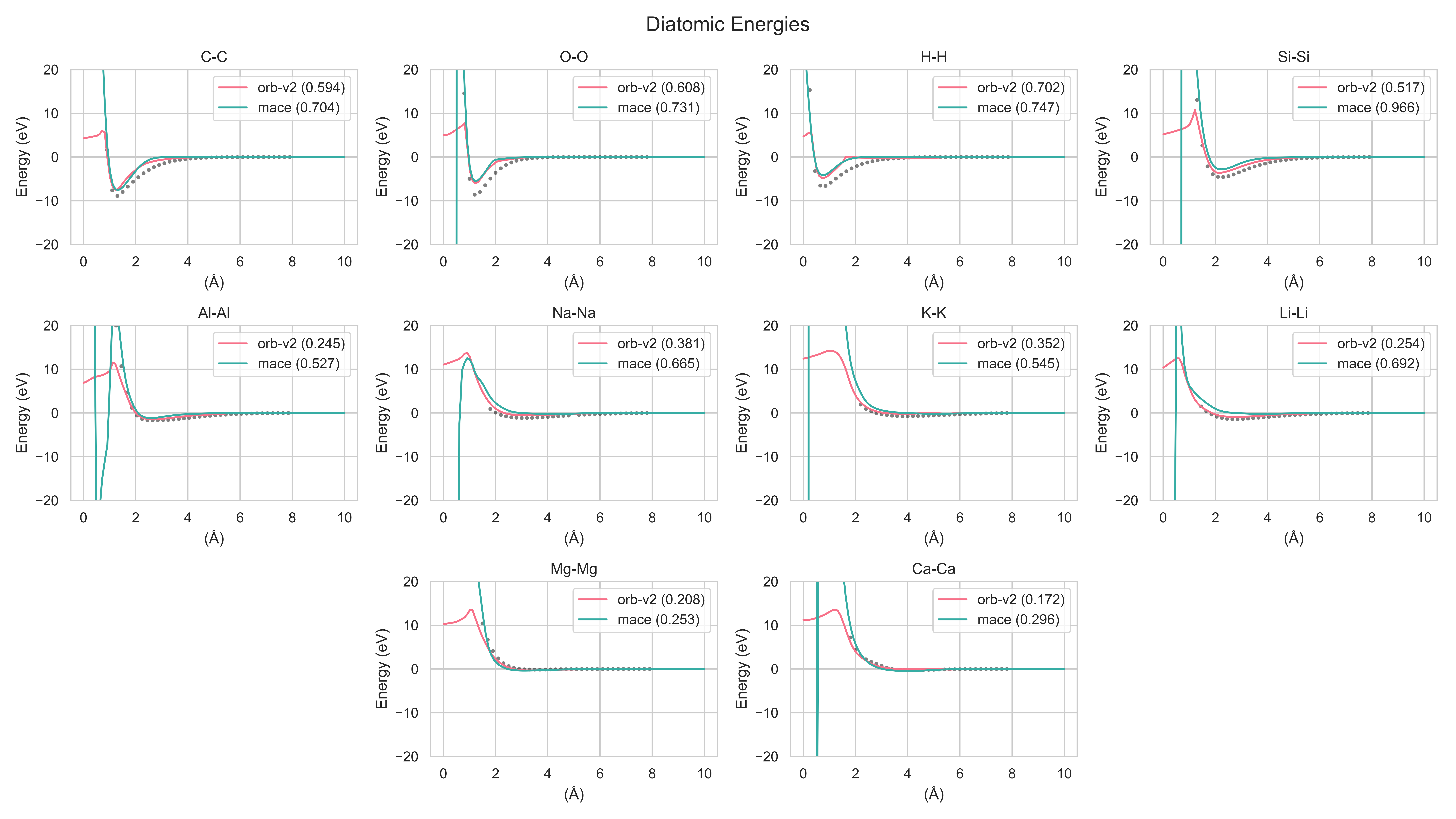}
    }
    \makebox[\textwidth][c]{
        \includegraphics[width=1.2\textwidth]{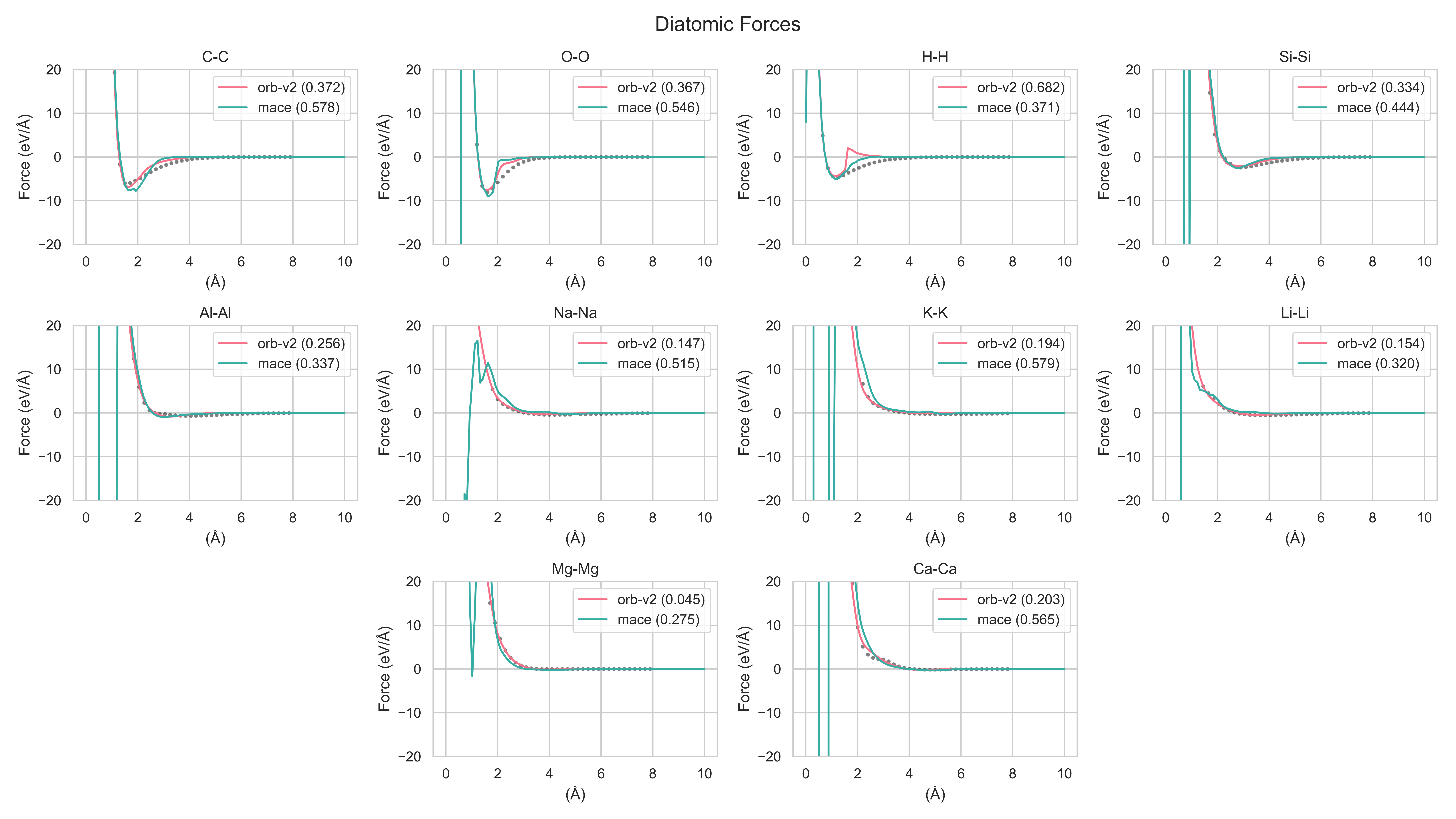}
    }
    \caption{Homo-Nuclear diatomic comparisons between MACE and Orb-v2 for a selection of elements (selected based on their common occurrence in Zeolite chemistry). Grey markers on the plots represent DFT energies and forces computed using VASP. Values in legend brackets correspond to MAE between each method and DFT at the marked points. For forces, we consider scalar forces in the \(x\) dimension, but this choice is arbitrary. Particularly of note are MACE's sharp instability at the boundary of covalent bond radii for many element pairs, demonstrating that equivariant models can also have failure modes with the potential to cause unstable simulations. Orb-v2 matches DFT accuracy more closely across the board, with the exception of H-H bond distances at the ~2\text{\AA} distance. }
    \label{fig:diatomics}
\end{figure}

\end{document}